\documentclass[10pt,aps,twocolumn,superscriptaddress,preprintnumbers,nofootinbib]{revtex4-1} 
\usepackage{amsmath,amssymb,graphicx,natbib,bm}
\usepackage{enumerate}
\usepackage{hyperref}
\usepackage{color}

\newcommand{\beq}{\begin {equation}}  
\newcommand{\eeq}{\end   {equation}} 
\newcommand{\bea}{\begin {eqnarray}} 
\newcommand{\eea}{\end   {eqnarray}}  
\newcommand{\baa}{\begin {array}   } 
\newcommand{\eaa}{\end   {array}   }     
\newcommand{\bit}{\begin {itemize} }
\newcommand{\eit}{\end   {itemize} }
\newcommand{\be }{\begin {equation}} 
\newcommand{\ee }{\end   {equation}}
\newcommand{\nn }{\nonumber        }

\newcommand{\tbox}[1]{\mbox{\tiny #1}}

\newcommand{\eff}{\textrm{eff}}

\begin{document}


\preprint{UTTG-10-14}
\preprint{TCC-011-14}

\title{Stabilizing Electroweak Vacuum in a Vector-like Fermion Model}

\author{Ming-Lei Xiao}
\email{mingleix@utexas.edu}
\affiliation{Theory Group, Department of Physics,\\The University of Texas at Austin,  Austin, TX 78712 U.S.A.}
\author{Jiang-Hao Yu}
\email{jhyu@utexas.edu}
\affiliation{Theory Group, Department of Physics,\\The University of Texas at Austin,  Austin, TX 78712 U.S.A.}
\affiliation{Texas Cosmology Center, \\The University of Texas at Austin,  Austin, TX 78712 U.S.A.}


\begin{abstract}
%
To avoid possible electroweak vacuum instability in the vector-like fermion model,
we introduce a new singlet scalar to the model, which couples
to the vector-like fermion, and also mixes with the Higgs boson after spontaneous symmetry breaking.
We investigate the vector-like fermion predominantly coupled to the third generation quarks,
and its mass is generated from the vacuum expectation value of the new scalar field in the model.
In this setup, as running towards high energies,
the new scalar provides positive contribution to the running 
of the Higgs quartic coupling,
and the matching on the scale of the scalar mass gives rise to a threshod effect that lifts up the Higgs quartic coupling strength.
The two effects help stabilize the electroweak vacuum of the Higgs potential.
Therefore, this setup could evade possible vacuum instability in the vector-like fermion model.
We show that  
a large range of parameter space is allowed to have both stable Higgs vacuum and 
perturbativity of all the running couplings, up to the Planck scale. 
We also examine the experimental constraints from the electroweak precision observables such as 
oblique corrections $S, T$ and non-oblique corrections to the $Zb_L\bar{b}_L$ coupling,  
the Higgs coupling precision measurements,
and the current LHC direct searches.
\end{abstract}

\maketitle


\section{Introduction}
\label{sec:intro}

The discovery of the Higgs-like scalar boson at the Large Hadron Collider (LHC) 
is the great triumph of the standard model (SM) of particle physics. 
The Higgs boson mass, 
was measured at the ATLAS and CMS with reasonable accuracy: $m_h = 125.9 \pm 0.4$ GeV~\cite{Beringer:1900zz}. 
Now that all the parameters of the SM are determined by experimental data,
the completion of the SM evoked our interest in its high energy behavior 
such as Higgs vacuum stability.
The measured value of the Higgs boson mass leads to a very intriguing situation. 
The most accurate analysis of the electroweak vacuum stability in the SM was performed in Ref.~\cite{Degrassi:2012ry, Buttazzo:2013uya}, showing that:
the theory sits near the boundary between stable phase and instable phase of the vacuum structure
if there is no new physics (NP) beyond the SM.
Therefore, 
NP should be introduced to stabilize the electroweak vacuum of the Higgs potential.

There are already many kinds of NP models avaliable to address the TeV scale physics. 
These models involve various extensions of the scalar sector, fermion sector,
and/or gauge boson sector of the SM. 
The Higgs vacuum stability serves as a criteria to justify the high energy 
behavior of these NP models~\cite{EliasMiro:2012ay}.
Among these, the models with extended fermion sector would
worsen the potential instability~\cite{Kobakhidze:2013pya, Khan:2012zw, Dev:2013ff, Rodejohann:2012px}. 
We will focus on  a vector-like fermion model 
predominantly coupled to the third generation quarks~\cite{Dawson:2012di, Aguilar-Saavedra:2013qpa, Cacciapaglia:2010vn,Dobrescu:2009vz,Berger:2012ec}, 
and manage to improve the stability of the electroweak vacuum by modifying its particle content. 
The vector-like fermion model is less constrained by the experimental data than
the models with extra chiral fermions.
Due to non-decoupling behavior of the chiral fermion,
the electroweak data from the precision measurements,
the Higgs coupling measurements, and the LHC direct searches
put very strong limits on its parameter space~\cite{Djouadi:2012ae,Eberhardt:2012gv}.
Moreover, due to the tight constraints on the light quarks from flavor physics, 
the heavy vector-like fermion can not significantly couple to the first two generations.

In order to find out the true vacuum state and analyze its stability, 
we need to investigate the Higgs effective potential. 
Since the instability occurs at energies much higher than the electroweak scale, 
the effective potential is well
approximated by the renormalization group (RG) improved tree-level expression
for the large field values  $h \gg v=246$ GeV~\cite{Buttazzo:2013uya}:
\bea
V_{\eff} (h) =\frac{\lambda(\mu)}{4} h^4,
\eea
where the Higgs quartic coupling $\lambda(\mu)$ runs with the renormalization scale $\mu$.
Generally speaking, if the Higgs quartic coupling $\lambda(\mu)$ becomes negative,
the effective potential becomes instable through developing a minimum much deeper than the realistic 
minimum.
As we know, in the SM the Higgs quartic coupling becomes negative at around $10^{10}$ GeV according to 
complete next-to-next-to-leading order (NNLO) calculation of the running Higgs quartic coupling~\cite{Degrassi:2012ry}. 
%
As shown in the Fig.~3 of the Ref.~\cite{Buttazzo:2013uya}, in which the 
measured uncertainties of the top quark and the Higgs masses are taken into account, 
the SM Higgs vacuum state lies in a narrow region of the metastable phase. 
In the vector-like fermion model, 
new vector-like fermion  
has negative contributions to the $\beta$ function 
of the running Higgs quartic coupling, which drives the Higgs potential 
towards the absolute instable phase. 

To cure the potential instability in the vector-like fermion model, 
we would like to 
modify the particle content and interactions in the model.
In the RG running, 
new bosonic particles provide positive contributions to 
the running of the Higgs quartic coupling, 
while new fermionic particles 
contribute negatively.
Here we suggest a very simple and economical extension 
by adding a new scalar singlet to the vector-like fermion model.
This new singlet scalar couples to the Higgs boson, 
and thus provides a positive contribution 
to the running of the Higgs quartic coupling.
However, if the new scalar only couples to the Higgs boson but not to others,
the quartic coupling of the new scalar will increase   
as the energy goes higher and higher.
Thus it is likely that the scalar quartic coupling by itself runs into non-perturbativity region 
at or below the Planck scale.
To avoid this possible problem, we require that the new scalar couples to 
the vector-like fermion, which provides a negative contribution to the running of the scalar quartic coupling, and thus controls the growth of the scalar quartic coupling strength at high energies. 
We assume that the new scalar singlet has a nonzero vacuum expectation value (vev).
To make the model more predictive, 
this vev also generates the mass term for the vector-like fermion.
In this setup, there are two effects to lift up the running Higgs quartic coupling strength. 
First, the new scalar provides positive contribution to the $\beta$ function of the running Higgs quartic coupling.
Second, in the matching on the scale of the scalar mass, the Higgs quartic coupling obtains a positive threshod shift~\cite{Lebedev:2012zw, EliasMiro:2012ay}.
Therefore, the new scalar could stabilize the electroweak vacuum of the Higgs potential, 
and this setup could evade possible vacuum instability in the vector-like fermion model.

The paper is organized as follows. 
In the next section we set up the Lagrangian of the model, and 
obtain the mass spectrum in the model. 
In Sec. 3, we present a study of vacuum stability through the one-loop renormalization group running. 
The matching between different energy scales are carefully treated.
We consider the theoretical bounds on the masses of the heavy particles from perturbative unitarity in Sec. 4.
In Sec. 5, the precision electroweak observables, including the oblique corrections $S$, $T$, and non-oblique corrections 
to the $Zb_L\bar{b}_L$ couplings, are examined.
In Sec. 6, we perform a global fit on the Higgs coupling precision measurements and put constraints on the parameter space.
Then we discuss the direct searches on the heavy scalar and the vector-like fermion.
Finally we summarize the constraints on the parameter spaces in the conclusion section.
In Appendix A we provide details of the relevant electroweak Lagrangian in the model. 
%
In Appendix B we list the one-loop renormalization group equations. 
Appendix C presents the detailed calculations on the oblique corrections.
In Appendix D we list the partial decay widths of the heavy particles.


\section{The Model}
\label{sec:model}

We consider an extension of the SM, by adding a vector-like fermion singlet $\psi$ with charge $+2/3$, 
and a real neutral singlet scalar $\chi$.
Since the vector-like fermion has the same quantum number as the right-handed up-type quarks, they will mix together. 
Due to the tight constraints on the up and charm quarks from flavor physics, 
we assume the vector-like fermion only mixes with the top quark as a fermionic top partner.  
The new scalar interacts with the Higgs doublet in the potential, 
which induces mixing between the scalar and the SM Higgs boson
 after spontanous symmetry breaking.

The scalar potential reads
\bea
	V(\Phi,\chi) &=& - \mu^2_H \Phi^\dagger \Phi + \lambda_H (\Phi^\dagger \Phi)^2 \nonumber\\
	&& -\frac{\mu^2_S}{2} \chi^2 + \frac{\lambda_S}{4} \chi^4 + \frac{\lambda_{SH}}{2} (\Phi^\dagger \Phi) \chi^2,
\eea
where $\Phi$ is the SM Higgs doublet
\bea
	\Phi =  \left( \begin{array}{c} \pi^+\\ \frac1{\sqrt2} ( \phi + i \pi^0) \end{array}\right).
\eea 
%
Requiring the scalar potential to be positive for asymptotically large values of the fields, we 
obtain the following conditions
\bea
	4\lambda_S \lambda_H > \lambda_{SH}^2, \nn\\
	\lambda_H > 0, \,\, \lambda_S > 0.
\eea
In general the scalar potential develops non-zero vacuum expectation values for both the Higgs and the singlet:
\bea
	\langle \Phi \rangle =  
	\left( \begin{array}{c} 0 \\  
	\frac{v}{\sqrt2} \end{array}\right), 
	\quad \langle \chi \rangle = u,
\eea
with the following relations from tadpole conditions
\bea
\mu_H^2 &=& \lambda_H v^2 + \frac{\lambda_{SH} u^2}{2},\\
\mu_S^2 &=& \lambda_S u^2 + \frac{\lambda_{SH} v^2}{2}.
\eea
If we assume the new physics is at the TeV scale, the new scale $u$ is larger than the electroweak scale $v$.
After symmetry breaking, there are mass mixing between the SM Higgs $\phi$ and the scalar $\chi$. 
The mixing matrix is
\bea
	{\mathcal M}^2_{S} = \left( \begin{array}{cc}
						2 \lambda_H v^2 & \lambda_{SH} v u\\
						\lambda_{SH} v u &2 \lambda_S u^2 
						\end{array}\right).
\eea
Diagonalizing the above matrix, we obtain the mass squared eigenvalues
\bea
	m^2_{h,S} &=& \lambda_H v^2 + \lambda_S u^2 \mp 
		  \sqrt{(\lambda_S u^2 - \lambda_H v^2)^2 +  \lambda_{SH}^2 u^2 v^2 }.\nn\\
		  \label{eq:scalarmass}
\eea
and the eigenvectors
\bea
	\left( \begin{array}{c} h \\ S \end{array}\right) = 
	\left( \begin{array}{cc}
	\cos\varphi & -\sin\varphi\\
	\sin\varphi & \cos\varphi
	\end{array}\right)
	\left( \begin{array}{c} \phi \\ \chi \end{array}\right), 
\eea
where the mixing angle  $\varphi$ given by
\bea
	\tan 2\varphi = \frac{\lambda_{SH} u v}{\lambda_S u^2 - \lambda_H v^2}.
	\label{eq:scalarangle}
\eea
The Eqs.~(\ref{eq:scalarmass}) and~(\ref{eq:scalarangle}) can be inverted to express 
the parameters $\lambda_H$, $\lambda_S$, and $\lambda_{SH}$ in terms of 
the physical quantities $m_{h,S}$ and the mixing angle $\varphi$:
\bea
	\lambda_H & = & \frac{m_{h}^2 \cos^2\varphi + m_{S}^2 \sin^2\varphi }{2 v^2}  ,\nn\\
	\lambda_S & = &  \frac{m_{S}^2 \cos^2\varphi + m_{h}^2 \sin^2\varphi }{2 u^2}  ,\nn\\
	\lambda_{SH} & = & \frac{m_{S}^2 - m_{h}^2}{2 uv} \sin 2\varphi. 
\eea

The general Yukawa couplings involving in the vector-like fermion $\psi$ and the top quark $u_3$ read 
\bea
	-{\mathcal L}_{\textrm{Yukawa}} & = &  
	\frac{y_M}{\sqrt2} \chi \overline{\psi}_L \psi_R +   \frac{\lambda_{T}}{\sqrt2} \chi \overline{\psi}_L u_{3R} \nn\\
	&&    + y_t \overline{Q}_L \tilde{H} u_{3R} + y_T  \overline{Q}_L \tilde{H} \psi_R + h.c.,
\eea 
where $\tilde{H} = i \sigma_2 H^\star$  and  $Q_L$ is the left-handed third-generation quark doublet 
$Q_L =  \left( \begin{array}{c}  u_{3L}\\ b_L \end{array}\right)$.
We can also write down a Dirac mass term of the vector-like fermion:
\bea
	-{\mathcal L}_{\textrm{mass}} = m_{D}  \overline{\psi}_L \psi_R + h.c.
\eea
After spontanous symmetry breaking, the vector-like fermion and the top quark mix together.
The mass mixing matrix between $(u_{3L}, \psi_L)$ and $(u_{3R}, \psi_R)$ is written as
\bea
	{\mathcal M}_{F} = \left( \begin{array}{cc}
						\frac{y_t v}{\sqrt2} & \frac{y_T v}{\sqrt2} \\
						\frac{\lambda_T u}{\sqrt2} & m_D + \frac{y_M u}{\sqrt2}
						\end{array}\right).
\eea
To make the model more predictive, we assume that the mass of the vector-like fermion are 
purely generated from the spontanous symmetry breaking, such that $m_D=0$.
Through a redefinition of the fields $(t_R, T_R)$, one can always rotate away 
one off-diagonal element of the mass matrix, such that $y_T = 0$, or $\lambda_T = 0$.
As is in the literature~\cite{Dawson:2012di}, we choose to rotate $(t_R, T_R)$ by an angle 
$\tan^{-1}(\lambda_T/(y_M))$ to have $\lambda_T = 0$. 
So after the rotation, the mass matrix becomes
\bea
	{\mathcal M}_{F} 
	= \left( \begin{array}{cc}
	   \frac{y_t v}{\sqrt2} & \frac{y_T v}{\sqrt2} \\
	   0 &  \frac{y_M u}{\sqrt2}
	   \end{array}\right).
\eea

To diagonalize the fermion mass matrix, we rotate the gauge eigenstates $(u_{3}, \psi)$ into 
the mass eigenstates $(t, T)$ using two $2\times 2$ unitary transformations 
\bea
	 \left(\begin{array}{c} t_{L,R} \\ T_{L,R} \end{array}\right) 
	 =  {\mathcal U}_{L,R}\left(\begin{array}{c} u_{3L,R} \\ \psi_{L,R} \end{array}\right) ,
\eea
where the unitary matrices are
\bea
	{\mathcal U}_{L,R} = 
	\left(\begin{array}{cc} \cos\theta_{L,R} & -\sin\theta_{L,R}  \\ 
	\sin\theta_{L,R} &  \cos\theta_{L,R} \end{array}\right) .
\eea
Thus, the mass matrix $\mathcal{M}_F$ transforms as
\bea
	{\mathcal U}_{L} \mathcal{M} {\mathcal U}_{R}^\dagger = \mathcal{M}_{\textrm{diag}} 
	= \left(\begin{array}{cc} m_t & 0 \\ 0 &  m_T \end{array}\right). 
\eea
The mass squared of the top quark $t$ and its partner $T$ are
\bea
	m_{t,T}^2 &=& \frac{1}{4}\left(y_t^2 v^2 + y_T^2 v^2 + y_M^2 u^2 \right) \nn\\
	&&\left[1 \mp \sqrt{1- \left(\frac{2 y_t y_M vu}{y_t^2 v^2 + y_T^2 v^2 + y_M^2 u^2 }\right)^2}\right].
	\label{eq:fermionmass}
\eea
and the mixing angles are
\bea
	\tan 2\theta_L  &=& \frac{2 y_T y_M v u}{y_M^2 u^2 - y_T^2 v^2 - y_t^2 v^2},\nn\\
	\tan 2\theta_R  &=& \frac{2 y_t y_T v^2}{y_M^2 u^2 + y_T^2 v^2 - y_t^2 v^2}.
	\label{eq:fermionmixing}
\eea
Note that the two mixing angles are not independent parameters, with the relation
\bea
	\tan \theta_R = \frac{m_t}{m_T}\tan \theta_L.
\eea
It is also useful to invert Eqs.~(\ref{eq:fermionmass}) and~(\ref{eq:fermionmixing}) 
to express the model parameters $(y_t, y_T, y_M)$ in terms of physical parameters $(m_t, m_T, \theta_L)$:
\bea
y_M     &=&\frac{\sqrt2 m_T}{u}\sqrt{\cos^2\theta_L+x_t^2\sin^2\theta_L},\nn\\
y_T     &=&\frac{\sqrt{2}m_T}{v}\frac{\sin\theta_L\cos\theta_L (1 - x_t^2)}{\sqrt{\cos^2\theta_L+x_t^2\sin^2\theta_L}},\nn\\
y_t     &=&\frac{\sqrt2 m_t}{v} \frac{1}{\sqrt{\cos^2\theta_L+x_t^2\sin^2\theta_L}},
\eea
where $x_t = m_t/m_T$.

In this model, 
the singlet scalar interacts with the SM particles 
in two ways: mixing with the SM Higgs boson, 
and interacting with the top quark through Yukawa coupling. 
In the top quark sector, the vector-like fermion mixes with the top quark.
The mass of the vector-like fermion is generated from the vevs of the symmetry breaking. 
This forces the masses of the two heavy fields $S$ and $T$ at the order of the scale $u$, which is assumed to be TeV scale. 
Let us summarize the new parameters in the model.
There are five independent parameters, which are chosen to be two heavy masses $m_S, m_T$, 
two mixing angles $ \varphi, \theta_L$, and the TeV symmetry breaking scale $u$.
In the following sections, we will use the short-hand notation for the mixing angles
\bea
	&& s_{\varphi}\equiv \sin\varphi,	\quad c_{\varphi}\equiv \cos\varphi, \nn\\
	&& s_{L}\equiv \sin\theta_L,	\quad c_{L}\equiv \cos\theta_L.	
\eea


\section{Vacuum Stability and Renormalization Group Equations}
\label{sec:vacuum}

In order to find out the true vacuum and investigate its stability, 
we should study the effective scalar potential which includes the radiative loop corrections and RG-improved parameters.
At the one-loop order, the effective scalar potential is~\cite{Coleman:1973jx}, in the Landau gauge,
\bea
	V_{\eff}(\Phi,\chi) &=& V(\Phi,\chi) + \frac{1}{64\pi^2} \sum_i (-1)^{2 s_i} (2 s_i + 1) \nn\\
	&& M_i^4(\Phi^2, \chi^2) \left[ \ln \frac{ M_i^2(\Phi^2, \chi^2)}{\mu^2} - c_i \right],
\eea
where $M_i^2(\Phi^2, \chi^2)$ are the field dependent mass-squared, and the index $i$ runs over all the fields in the model.
Here $c_i$ are constants that depend on the renormalization scheme. 
We choose the $\overline{\textrm{MS}}$ scheme, with $c_i = 3/2$ for scalars and fermions, and $c_i = 5/6$ for vector bosons.
The effective scalar potential $V_{\eff}$ must develop a realistic minimum at the electroweak scale $v$, corresponding to the SM vev. 
The stability condition on the Higgs vacuum is dependent on the behavior of $V_{\eff}$ in the large-field limit $h \gg v=246$ GeV.
This condition is essentially equivalent to the requirement~\cite{Degrassi:2012ry} that the Higgs quartic coupling $\lambda(\mu)$ never becomes negative below the Planck scale.
We will study the RGE running behavior of the Higgs quartic coupling $\lambda(\mu)$ in the  $\overline{\textrm{MS}}$ scheme. 

This requires us to work in the effective field theory framework, by integrating heavy particles out at their mass thresholds 
and matching all the running couplings between effective theories at different scales. 
At the scale of the scalar pole mass  $M_S$,  
we can integrate out the scalar singlet in the tree level potential $V(\Phi,\chi)$ 
using its equation of motion:
\bea
	\chi^2 = u^2 - \frac{\lambda_{SH}}{\lambda_S} (\Phi^\dagger \Phi - v^2/2).
\eea
Inserting the above equation back to $V(\Phi,\chi)$, 
we obtain the tree-level effective Higgs potential below the heavy mass threshold:
\bea
	V(\Phi) = \lambda^{\textrm{SM}}  (\Phi^\dagger \Phi - v^2/2 )^2,
\eea
 where 
 \bea
 	\lambda^{\textrm{SM}}  = \lambda_H - \frac{\lambda_{SH}^2}{4\lambda_S}.
	\label{eq:threshold}
 \eea
This shows that there is a tree-level shift when we match the Higgs quartic  coupling $\lambda_H$ in the model 
to the  Higgs quartic coupling $\lambda^{SM}$ in the low energy effective theory.
This is consistent with the expression of the Higgs boson mass 
in the limit of $v \ll u$:
\bea
	m_h^2 = 2 v^2 \left(\lambda_H - \frac{\lambda_{SH}^2}{4\lambda_S}+ {\mathcal O}(v^2/u^2)\right).
\eea 
At the scale of the heavy fermion pole mass $M_T$, we also integrate out the heavy fermion using its equation of motion. 
The tree-level matching between the model and the low energy effective theory in the Yukawa sector tells us 
\bea
	y^{\textrm{SM}}_t = y_t - \frac{\lambda_T y_T}{y_M}.
\eea
Since we already take $\lambda_T = 0$ after the redefinition of the Yukawa couplings, no matching is needed for $y_t$.

Depending on different particle content in effective theories, there are different RGE running behaviors in different energy regions.

{\it Region I: scale $\mu < M_T, M_S$.} 
In this region, after integrating out all the heavy particles, we recover the SM as the low energy theory.
The SM one-loop RGE for the Higgs quartic coupling is
\bea
\frac{d\lambda}{d\ln \mu^2}&=& \beta_\lambda^{\textrm{SM}} = \frac{\lambda}{(4\pi)^2} \bigg[ 12 \lambda +6 y_t^2
-\frac{9 g_1^2}{10}-\frac{9 g_2^2}{2}\bigg]  \nn\\
&+& \frac{1}{(4\pi)^2} \bigg[ -3 y_t^4 +\frac{9 g_2^4}{16}+\frac{27 g_1^4}{400}+\frac{9 g_2^2 g_1^2}{40} \bigg],
\eea
where $g_1 = \sqrt{5/3} g_Y$ is the hypercharge gauge coupling in GUT normalisation, 
and $g_2$ the weak $SU(2)_L$ gauge coupling.
Note that for a light Higgs, the running behavior is mainly controlled by the top quark Yukawa coupling, 
which drives $\lambda$ towards more negative values. 
If there is no new particle running in the loop, $\lambda$ would eventually become negative at high energy scale
 around $10^{10}$ GeV.

In order to determine the boundary condition for $\lambda(\mu)$ at the renormalization scale $\mu$, 
one need to know 
how the $\overline{\textrm{MS}}$ renormalized Higgs quartic coupling $\lambda (\mu)$ relates to the SM input parameters. 
Here the SM input parameters are taken to be the SM pole masses $M_h, M_t, M_W, M_Z$ and Fermi constant $G_F$, $\alpha_s(M_Z)$.
The relation that connects $\lambda (\mu)$ to the SM input parameters can be written as
\bea
	\lambda^{\textrm{SM}} (\mu) = \frac{G_F M_h^2}{\sqrt{2}}\left[1 + \Delta_h (\mu)\right],
\eea 
where $\Delta_h (\mu)$ represents electroweak one loop radiative corrections at the scale $\mu$~\cite{Sirlin:1985ux}. 
Similarly, the boundary condition for $y_t$ can be determined from the relation between the pole mass and its running mass:
\bea
	y_t^{\textrm{SM}} (\mu) =  (\sqrt{2} G_F)^\frac12 M_t \left[1 + \Delta_t (\mu)\right],
\eea 
where $\Delta_t (\mu)$ denotes the electroweak radiative corrections~\cite{Hempfling:1994ar}.
In the RGE running, we start from the scale of the top pole mass $M_t$. 
The boundary conditions of the couplings at the $M_t$ scale are taken from the two-loop matched values 
presented in Ref.~\cite{Buttazzo:2013uya}.

{\it Region II (a): scale $\mu \geq M_T$ and $\mu < M_S$.} 
There are two cases in the intermediate region since the model could have either $M_T < M_S$ or $M_T > M_S$. 
Let us first discuss the case (a): $M_T < M_S$. 
Due to $\lambda_T = 0$, there is no matching condtion on the top-Yukawa coupling at the scale of the heavy fermion mass $M_T$.
Above this scale, the heavy fermion contributes to the one-loop running of
the gauge couplings, Yukawa couplings, and Higgs quartic coupling $\lambda$.
The RGE for the Higgs quartic coupling becomes
\bea
&&\frac{d\lambda}{d\ln\mu^2} = \beta_\lambda^{\textrm{SM}} + \frac{1}{(4\pi)^2} 
\bigg[   6 \lambda  y_T^2 - 3 y_T^4 -   6 y_T^2 y_t^2 \bigg].
\eea
%
%
Due to the negative contributions from the additional terms in the above RGE, 
we expect the scale, 
at which the Higgs quartic coupling becomes negative, to be lower than that in the SM.
Therefore, in the pure vector-like fermion model, the Higgs vacuum instability problem is worse than that in the SM. 

{\it Region II (b): scale $\mu \geq M_S$ and $\mu < M_T$.} 
This is the case $M_T > M_S$. 
%
According to Eq.~\ref{eq:threshold}, 
the Higgs quartic coupling receives a positive shift at the $M_S$ threshold
 \bea
 	\lambda_H   =  \lambda^{\textrm{SM}} + \frac{\lambda_{SH}^2}{4\lambda_S},
	\label{eq:threshold2}
 \eea
which is the matching condition on the Higgs quartic coupling. 
The heavy scalar also changes the RGE running behavior of the Higgs quartic coupling, which becomes
\bea
\frac{d\lambda_H}{d\ln\mu^2}&=&  \beta_\lambda^{\textrm{SM}} +  \frac{1}{(4\pi)^2}  \frac14 \lambda_{SH}^2.
\label{eq:rgeIIb}
\eea
The positive contribution from the last term in the Eq.~\ref{eq:rgeIIb} 
delays the occurrence of the vacuum instability at high energies.
Thereforth, the above two effects on the RGE running could avoid the possible vacuum instability.

However, we have to worry about the perturbativity bounds on the scalar quartic coupling $\lambda_S$. The RGE running of the scalar coupling is
\bea
\frac{d\lambda_S}{d\ln\mu^2}&=& \frac{1}{(4\pi)^2} \bigg[ 9 \lambda_S^2 + \lambda_{SH}^2\bigg].
\eea
If there is no new heavy fermion coupled to the heavy scalar, 
it is very likely that the scalar coupling $\lambda_S$ blows up at some energy scale, and thus voilate the pertubativity bounds.
Although adding a heavy scalar could solve the instability problem, there is another problem from 
the pertubativity bounds on the scalar coupling 
in the model with only scalar sector extension.

{\it Region III: scale $\mu \geq M_T, M_S$.} 
In this region, both the heavy fermion and the heavy scalar are involved in the RGE running. 
%
If $M_T < M_S$, the quartic Higgs coupling would receive the same tree level threshold correction 
as the Eq.~\ref{eq:threshold2} 
at the boundary of the Region III. 
The full RGE running of the Higgs quartic coupling is 
\bea
\frac{d\lambda_H}{d\ln\mu^2}&=& \beta_\lambda^{\textrm{SM}} +  \frac{1}{(4\pi)^2} \bigg[   6 \lambda_H y_T^2  + \frac14 \lambda_{SH}^2 -3 y_T^4   - 6y_t^2 y_T^2   \bigg].\nn\\
\eea
We notice that in above RGE the negative contribution to $\beta$ function of the Higgs quartic coupling from $y_T$ is vastly softened by the positive contribution from $\lambda_{SH}$, especially at high energies.
Including the positive threshold shift, the vacuum instability problem could be evaded.
On the other hand, we need to take care of the perturbativity bounds on the scalar coupling.
The RGE running of the scalar coupling $\lambda_S$ is
\bea
\frac{d\lambda_S}{d\ln\mu^2}&=& \frac{1}{(4\pi)^2} \bigg[ 9 \lambda_S^2 + \lambda_{SH}^2 + 6 y_M^2 \lambda_S - 3 y_M^4  \bigg].
\eea
%
%
Without the $y_M^4$ term in the above equation, $\lambda_S$ could blow up and reach the Landau pole at some scale. 
The presence of new Yukawa coupling plays a role to avoid this trouble.
%

In our numerical scan, we requires that at all the running scales below Planck scale, 
\bea
	\lambda_H(\mu) > 0, \quad 0 < \lambda_S(\mu) < 4 \pi.
	\label{eq:condition1}
\eea
The evolution of $\lambda_{SH}$ is written as
\bea
\frac{d\lambda_{SH}}{d\ln\mu^2}&=& \frac{1}{(4\pi)^2} \bigg[  \lambda_{SH} \bigg( 2\lambda_{SH}  + 6 \lambda_H +  3 \lambda_S 
 + 3 y_t^2 \nn\\
&&  +3 y_T^2 +  3 y_M^2 -\frac{9 g_1^2}{20}-\frac{9 g_2^2}{4}\bigg)   - 6  y_T^2y_M^2 \bigg].
\eea
Since $\lambda_{SH}$ could be either positive or negative, we only require  
\bea
	|\lambda_{SH}(\mu)| < 4 \pi.
	\label{eq:condition2}
\eea
The RGE running of the new Yukawa couplings are  
\bea
\frac{dy_T^2}{d\ln\mu^2}&=&\frac{y_T^2}{(4\pi)^2} \bigg[ \frac{9}{2} y_T^2 + \frac92 y_t^2  \nn\\ 
&& +\frac14 y_M^2  
-\frac{17}{20}g_1^2 - \frac{9}{4}g_2^2 - 8 g_3^2) \bigg], \\
\frac{dy_M^2}{d\ln\mu^2}&=&\frac{y_M^2}{(4\pi)^2} \bigg[ y_T^2  + \frac{9}{2} y_M^2
-\frac{8}{5}g_1^2 - 8 g_3^2  \bigg].
\eea
We require all the Yukawa couplings to be in the perturbative region at all energies below the Planck scale.

\begin{figure}[t]
\begin{center}
\includegraphics[width=0.49\textwidth]{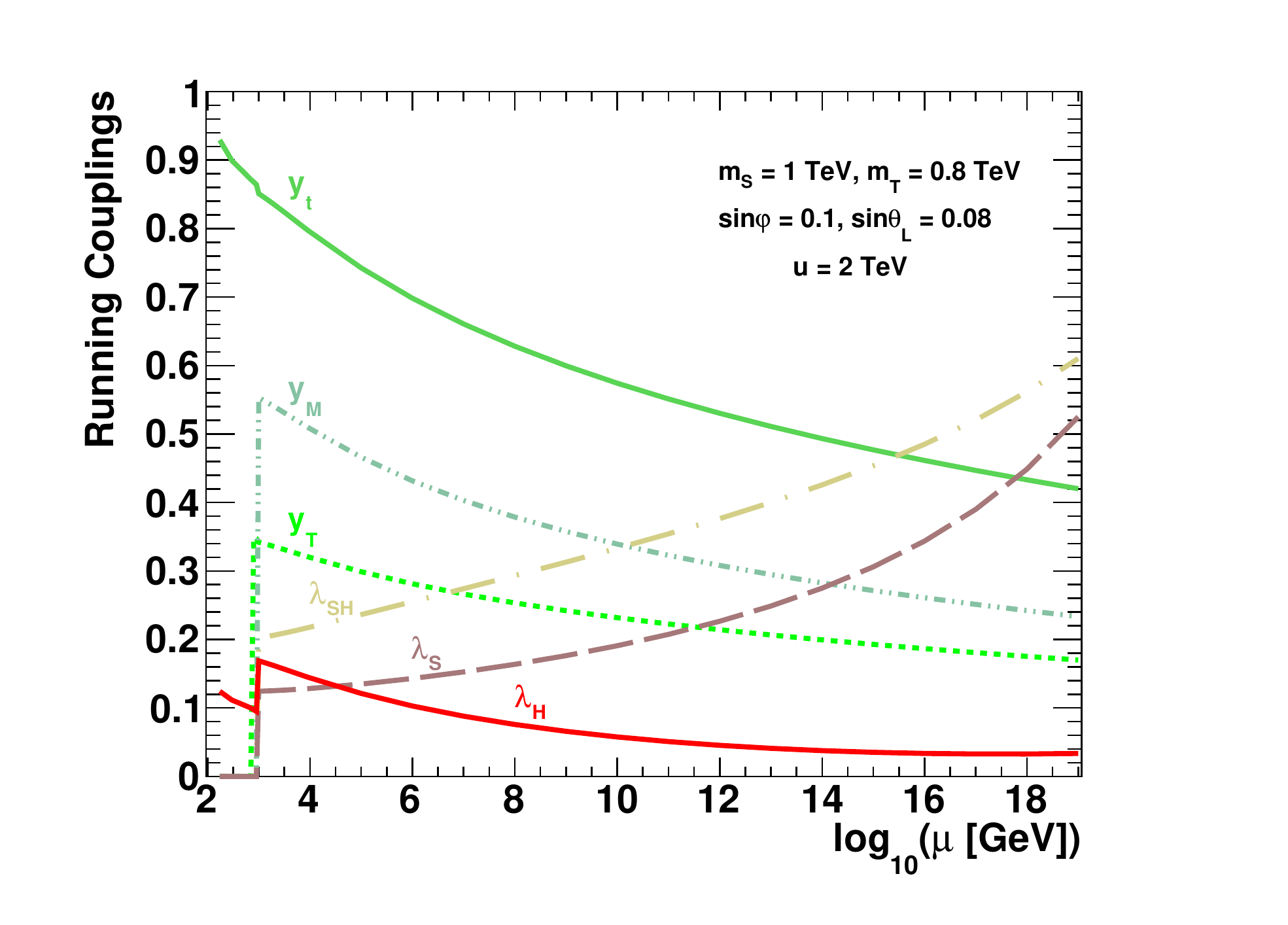} 
\caption{\small The RGE running of the Yukawa and scalar couplings in the model. All parameters are defined in the $\overline{\textrm{MS}}$ scheme. The starting point of the running is $m_t(M_t)$. The benchmark point:  
$m_S = 1$ TeV, $m_T = 800$ GeV, $\sin\varphi = 0.1$, $\sin\theta_L = 0.08$ and $u = 2$ TeV, is taken.}
\label{fig:benchrge}
\end{center}
\end{figure}

\begin{figure}[t]
\begin{center}
\includegraphics[width=0.49\textwidth]{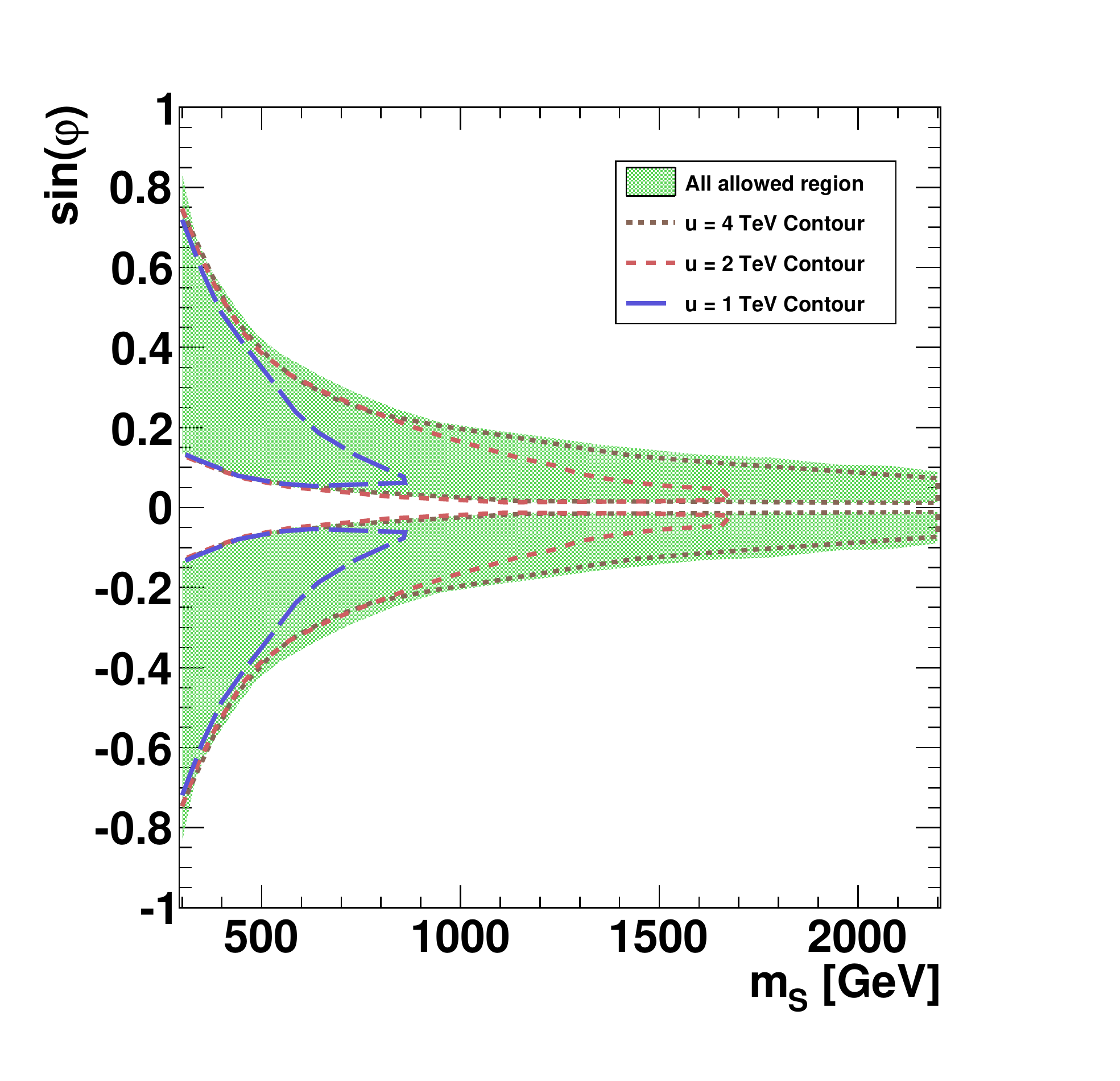} 
\caption{\small The allowed parameter region of 
the scalar mass and mixing angle   $(m_S, s_\varphi)$ satisfying the vacuum stability of the Higgs potential 
and perturbativity of all the running couplings. 
The dashed lines are the allowed contours $(m_S, s_\varphi)$ for different fixed scale $u$.}
\label{fig:rgems}
\end{center}
\end{figure}

\begin{figure}[t]
\begin{center}
\includegraphics[width=0.49\textwidth]{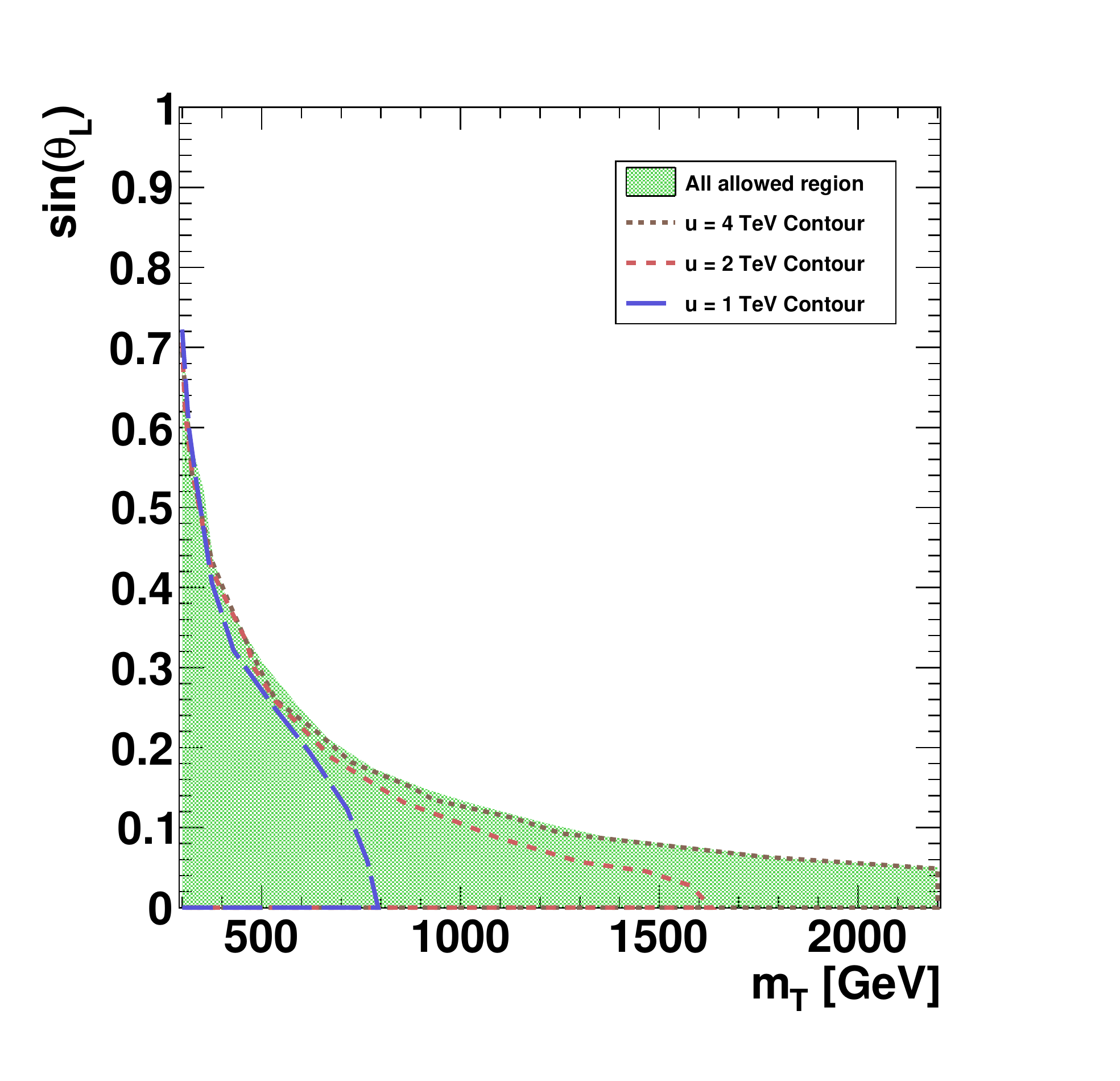} 
\caption{\small The allowed parameter region of   
the vector-like fermion mass and mixing angle  $(m_T, s_L)$ satisfying the vacuum stability of the Higgs potential 
and perturbativity of all the running couplings. 
The dashed lines are the allowed contours $(m_T, s_L)$ for different fixed scale $u$.}
\label{fig:rgemf}
\end{center}
\end{figure}

The complete RGE in three regions are listed in the Appendix B. 
To illustrate, we show the RGE running of the Yukawa and scalar couplings  in Fig.~\ref{fig:benchrge} 
for a typical parameter point: 
$m_S = 1$ TeV, $m_T = 800$ GeV, $\sin\varphi = 0.1$, $\sin\theta_L = 0.08$ and $u = 2$ TeV
. 
Using the Eq.~\ref{eq:condition1} and Eq.~\ref{eq:condition2},
we could put constraints on the parameter space in the model.
So we perform a numerical scan over a large range of the parameter space for all parameters: the masses $m_S, m_T$ 
and the mixing angles $s_\varphi, s_L$, and the scale $u$. 
Fig.~\ref{fig:rgems} and Fig.~\ref{fig:rgemf} show the allowed parameter space satisfying the stability
and the perturbativity conditions. 
As expected, if the mixing angle $s_\varphi$ is too small, and the scalar mass is light,
the scalar can not give enough lift on the Higgs quartic coupling.
In this small parameter region 
the Higgs quartic coupling will become negative below the Planck scale.
%
%
%
%
So in Fig.~\ref{fig:rgems} 
we notice there is a small region 
where the Higgs vacuum is instable.
Fig.~\ref{fig:rgems} also shows the zero $s_\varphi$ is always excluded. 
This indicates that it is not allowed to take the decoupling limit 
in the scalar sector. 
On the other hand, the parameter region where the scalar and the Higgs have a large mixing
is disfavored, especially when the scalar is heavy. 
The reason for this is that
the scalar quartic couplings will increase as evoluting to the high energy scale, and 
eventually become nonperturbative. 
%
Indeed, Fig.~\ref{fig:rgems} shows the region with large mixing angle $s_\varphi$ is excluded.
If we fix the scale $u$ (dashed contours in Fig.~\ref{fig:rgems}), there is a strict bound on the mass of the scalar
from the perturbativity limit on the scalar coupling strength $\lambda_S$.
%
%
Regarding to the parameter space for the heavy fermion,
we expect that small $s_L$ is favored,
since the small mixing angle usually gives rise to small Yukawa couplings $y_T$. 
Small $s_L$ could keep the Higgs quartic coupling positive up to the Planck scale.
Fig.~\ref{fig:rgemf} exhibits this feature. 
As also shown in Fig.~\ref{fig:rgemf}, if we fix the scale $u$ (dashed contours), 
the mass of the vector-like fermion also has an upper bound since small Yukawa coupling $y_M$ is favored.
Finally, we notice that the Fig.~\ref{fig:rgems} is symmetric for the positive and negative value of the $s_\varphi$.
From now on, we will present the parameter space with only positive half of the whole range of $s_\varphi$.
%


\section{Perturbative Unitarity}
\label{sec:unitarity}

Although there is no bad $s$-dependent high energy behavior in the model, 
the tree-level perturbative unitarity could  put constraints on the masses and couplings of the heavy particles.
In the partial wave treatment~\cite{Lee:1977eg}, 
given the tree-level scattering amplitude ${\mathcal M} (s, \theta)$ of all possible $2\to 2$ scattering processes,
the partial wave amplitude with angular momentum $J$ is written as
\bea
	a_{J} = \frac{1}{32\pi} \int^{1}_{-1} d\cos\theta \, P_J(\cos\theta) {\mathcal M} (s, \theta),
\eea
where $s$ and $\theta$ are the total energy squared, and the scattering polar angle in the center of mass frame, respectively.
$P_J(\cos\theta)$ is the Legendre Polynomial. 
The unitarity  requires the following condition~\cite{Lee:1977eg,Chanowitz:1985hj,Dicus:2004rg}
\bea
	|{\textrm Re}(a_J)| \le \frac{1}{2}. 
	\label{eq:unit}
\eea

In the high energy limit, following the equivalent theorem~\cite{Cornwall:1974km,Yao:1988aj,Bagger:1989fc,He:1992nga},
the unitarity condition could be obtained by calculating the partial wave amplitudes 
of the coupled channels in the scalar sector.
It has been shown~\cite{Pruna:2013bma} that 
the dominant contribution in the coupled channels is the process $SS \to SS$. 
In the high energy limit,  tree level amplitude of the $SS \to SS$ is 
\bea
\mathcal{M}(\frac{1}{\sqrt2} SS\rightarrow \frac{1}{\sqrt2} SS)
&=& \frac{1}{64 v^2 u^2}\Big[
    6 (m_H^2 + 5 m_S^2) (v^2 + u^2)\nn\\
&+& 3 (m_H^2 + 15 m_S^2) (v^2 - u^2) \cos(2\varphi) \nn\\
&-& 6 (m_H^2 - 3 m_S^2) (v^2 + u^2) \cos(4 \varphi) \nn\\
&-& 3 (m_H^2 - m_S^2) (v^2 - u^2) \cos(6 \varphi)   \nn\\
&-& 12(m_H^2 - m_S^2) 2vu \sin^3(2\varphi)\Big]. 
\eea
Put it back to the unitarity condition Eq.~\ref{eq:unit}, we obtain the constraints on the parameter space. 
In the limit of no mixing between the Higgs and the scalar,  it gives a constraint on $m_S$ against $u$:
\begin{equation}
m_S<\sqrt{\frac{4\pi}{3}}u .
\end{equation}

On the other hand, the heavy fermion also has an upper bound on its mass and coupling $s_L$
from the requirement of the perturbative unitarity through the fermion anti-fermion scattering process.
At high energy $\sqrt{s} \gg m_T$, the tree level amplitude 
of the process 
$T  \bar{T} \to T  \bar{T} $ is 
\begin{align}
\mathcal{M}(T\bar{T}\rightarrow T\bar{T})_{\lambda_i \lambda_f}&
= m_T^2(u^{-2}c_L^4+v^{-2}s_L^4) \begin{pmatrix}
1	&	0	&	0	&	0\\
0	&	0	&	-1	&	0\\
0	&	-1	&	0	&	0\\
0	&	0	&	0	&	1
\end{pmatrix}
\end{align}
where $\lambda_i $ and $\lambda_f$ are the helicity states of the initial and final states.
$\lambda_i $ and $\lambda_f$ are taken to be one of the following helicity states $\{++,+-,-+,--\}$.
Diagonalizing it and taking the largest $s$ wave component, we have the unitarity condition
\begin{equation}
a_0^{max}=\frac{1}{16\pi}\left[m_T^2(u^{-2}c_L^4+v^{-2}s_L^4)\right]< \frac12.
\end{equation}
Similarly,  if there is  no mixing between the vector-like fermion and the top quark,  
it gives a constraint on $m_T$ against $u$:
\begin{equation}
m_T<\sqrt{8 \pi}u .
\end{equation}


\section{Precision Electroweak Measurements}
\label{sec:stzbb}

The presence of the new scalar $S$ and the vector-like fermion $T$ renders both
modified SM couplings and new electroweak couplings.
We summarize the relevant Lagrangian involving gauge couplings and the Higgs couplings in the appendix A.
These electroweak couplings have impact on the electroweak observables, precisely measured at the LEP and SLC.

The dominant NP effects on the electroweak observables 
appear in the gauge boson vaccuum polarization correlations, named oblique corrections~\cite{Peskin:1991sw},
parametrized by three independent parameters S, T and U: 
\bea
\label{eq:ewS}\alpha S &\equiv& 4e^2 \left [ \Pi^\prime_{33}(0)-\Pi^\prime_{3Q}(0)\right], \\
\label{eq:ewT}\alpha T &\equiv& \frac{e^2}{s_W^2 c_W^2 m_Z^2}\left [\Pi_{11}(0) - \Pi_{33}(0) \right], \\
\label{eq:ewU}\alpha U &\equiv& 4e^2 \left [ \Pi^\prime_{11}(0)-\Pi^\prime_{33}(0)\right].
\eea
where the notation $\Pi_{XY}$ with $X, Y = 1, 3, Q$
denotes the vacuum polarization amplitudes 
and 
$\Pi^\prime_{XY}(q^2) = \frac{d}{d q^2}\Pi_{XY}(q^2)$.
From the global fit of the  electroweak precision data,
the constraints on the S, T and U parameters can be obtained.
The following fit results are determined from the GFitter fit~\cite{Baak:2012kk} 
for the reference SM parameters $m_t = 173$ GeV and $m_h = 126$ GeV.
In the NP model, the contribution of the U parameter is usually very small and can be neglected. 
Fixing $U = 0$, the GFitter global fit results in
\bea
	\Delta S  &=& S^{\textrm{NP}} - S^{\textrm{SM}} = 0.05 \pm 0.09\\
	\Delta T  &=& T^{\textrm{NP}} - T^{\textrm{SM}} = 0.08 \pm 0.07.
\eea
and the correlation coefficient is taken to be $0.91$.

\begin{figure}[t]
\includegraphics[width=0.45\textwidth]{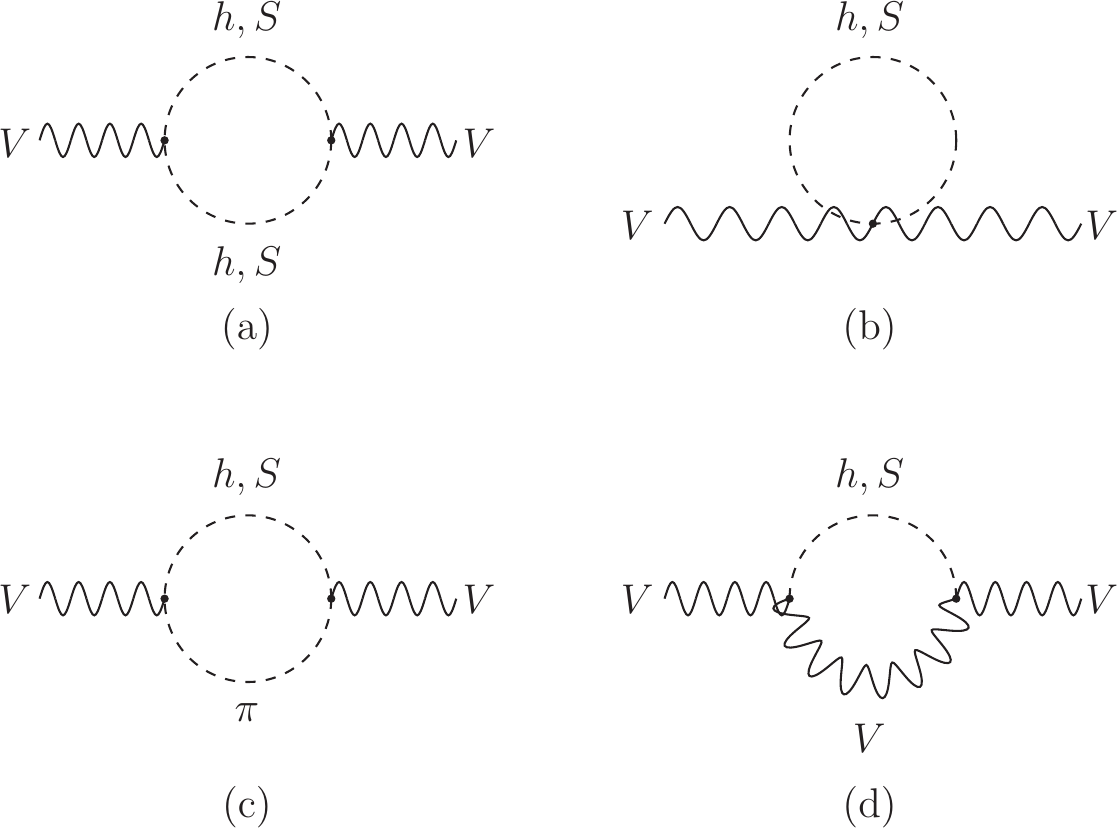} 
\caption{\small The one-loop Feynman diagrams of the vector boson self-energy $\Pi_{VV}$ due to the scalars in the loop. }
\label{fig:st_ss}
\end{figure}

\begin{figure}[t]
\includegraphics[width=0.45\textwidth]{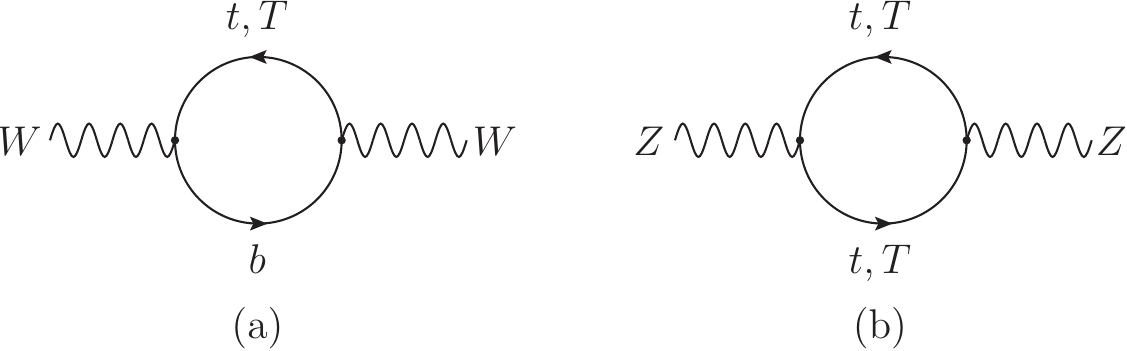} 
\caption{\small The one-loop Feynman diagrams of the vector boson self-energy $\Pi_{VV}$ due to heavy fermions in the loop. }
\label{fig:st_ff}
\end{figure}

We split the calculation on the oblique parameters into boson-loop contributions $T_S,\ S_S$ and 
fermion-loop contributions $T_F,\ S_F$, and consider them separately.
For the boson-loop contributions, the NP effect is only involved in the vacuum polarization amplitudes 
 where the Higgs or the heavy scalar are in the loop.
%
This is shown by Feynman diagrams in Fig.~\ref{fig:st_ss}.
%
Using the vector boson self-energy $\Pi_{VV}$ defined in the Appendix C, we obtain 
\bea
\Delta T_S &=&  s_{\varphi}^2 \Big[ T_s(m_S^2) - T_s(m_h^2)  \Big], \\
\Delta S_S &=&  s_{\varphi}^2 \Big[ S_s(m_S^2) - S_s(m_h^2)  \Big],
\eea
where the functions are defined as
\bea
\label{eq:oblqf}
T_s(m)&=&-\frac{3}{16\pi c_W^2}\bigg[\frac{1}{(m^2-m_Z^2)(m^2-m_W^2)}\notag\\
&&\times\Big(m^4\ln m^2-s_W^{-2}(m^2-m_W^2)m_Z^2\ln m_Z^2\notag\\
&+&s_W^{-2}c_W^2(m^2-m_Z^2)m_W^2\ln m_W^2\Big)-\frac{5}{6}\bigg], \\
S_s(m)&=&\frac{1}{12\pi}\bigg[\ln m^2-\frac{(4m^2+6m_Z^2)m_Z^2}{(m^2-m_Z^2)^2}\notag\\
&+&\frac{(9m^2+m_Z^2)m_Z^4}{(m^2-m_Z^2)^3}\ln\frac{m^2}{m_Z^2}-\frac{5}{6}\bigg].
\eea
%
%
%
%
%
Similarly, it is straightforward to calculate the oblique corrections 
due to the top quark and the vector-like fermion shown in Fig.~\ref{fig:st_ff}.
Subtracting the SM contributions due to the third generation quarks
\bea
	T^{\textrm{SM}}_F &=& \frac{3 m_t^2}{4\pi e^2 v^2} , \\
	S^{\textrm{SM}}_F &=& \frac1{2\pi} \left( 1 - \frac13 \log\frac{m_t^2}{m_b^2} \right) ,
\eea
we arrive at the final expressions
\bea
\label{eq:deltaTf}
\Delta T_F&=&T^{\textrm{SM}}s_L^2\Big[-(1+c_L^2) + s_L^2\frac{m_T^2}{m_t^2} \nn\\
&+& c_L^2\frac{2m_T^2}{m_T^2-m_t^2}\ln\frac{m_T^2}{m_t^2}\Big],\\
\Delta S_F&=&-\frac{s_L^2}{6\pi}\Big[(1-3c_L^2)\ln\frac{m_T^2}{m_t^2}+5c_L^2\nn\\
&-&\frac{6c_L^2m_t^4}{(m_T^2-m_t^2)^2}\Big(\frac{2m_T^2}{m_t^2}-\frac{3m_T^2-m_t^2}{m_T^2-m_t^2}\ln\frac{m_T^2}{m_t^2}\Big)\Big],
\eea
which agree with the results in Ref.~\cite{Dawson:2012di}.
%

\begin{figure}[t]
\begin{center}
\includegraphics[width=0.2\textwidth]{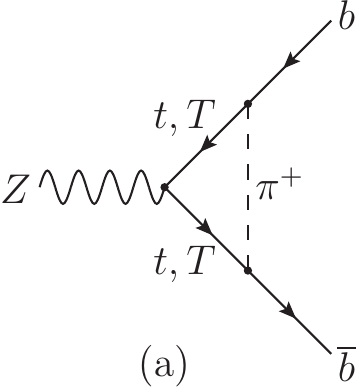} 
\includegraphics[width=0.2\textwidth]{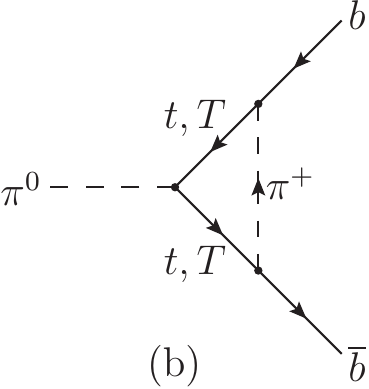} 
\caption{\small (a) the dominant one-loop Feynman diagrams in the t'Hooft-Feynman gauge;
(b) the only Feynman diagrams after the gaugeless limit is taken in the model. }
\label{fig:zbb}
\end{center}
\end{figure}

The only important non-oblique correction 
comes from the vertex correction of the $Zb\bar{b}$ coupling. 
In general, the effective  $Zb\bar{b}$ vertex can be parametrized as
\bea
	\frac{g_2}{c_W} \bar{b} \gamma^\mu \left[ g_{L} \frac{1-\gamma_5}2 + g_{R} \frac{1+\gamma_5}2 \right] b Z_\mu,
\eea
where 
\bea
	g_{L} &=&  g_{L}^{\textrm{SM}} + \delta g_{L}^{\tbox{NP}},\\
	g_{R} &=&  g_{R}^{\textrm{SM}} + \delta g_{R}^{\tbox{NP}}.
\eea
Here $g^{\textrm{SM}}$ denotes the SM coupling with radiative correction included, 
and $\delta g^{\tbox{NP}} $ represents the correction purely from the NP model.
In the SM, taking the leading $m_t$-dependent radiative corrections into account,
the SM couplings are
\bea
	g_{L}^{\textrm{SM}} &=& -\frac12 + \frac13 s_W^2 + \frac{m_t^2}{16\pi^2 v^2}, \\
	g_{R}^{\textrm{SM}} &=& \frac13 s_W^2.
\eea
In our model, there is no tree-level correction to the $Zb\bar{b}$ coupling.
However, at one-loop, flavor-dependent vertex corrections arise, and contribute to the $Zb_L\bar{b}_L$ coupling.
Fig.~\ref{fig:zbb}(a) shows the dominant one-loop Feynman diagram in the t'Hooft-Feynman gauge,  
in which the vector-like fermion and the top quark appear in the loop.
The presence of vertex corrections 
gives rise to non-zero 
$\delta g_L^{\tbox{NP}}$. 
To extract out the leading $m_T$-dependent terms explicitly, 
we perform the loop calculation in the ``gaugeless" limit~\cite{Barbieri:1992dq, Fleischer:1994cb, Abe:2009ni, Foadi:2011ip}, in which 
the $Z$ boson is treated as a non-propagating external field 
coupled to the current $J^\mu = \bar{b}_L \gamma^\mu b_L$. 
Using the Wald identity~\cite{Barbieri:1992dq, Fleischer:1994cb}, the leading contribution to the $Zb_L\bar{b}_L$ coupling 
can be obtained via the calculation of the higher dimensional operator
$\frac{\partial_\mu \pi^0}{m_Z} \bar{b}_L \gamma^\mu b_L$ , 
where $\pi^0$ is the Goldstone boson eaten by the $Z$ boson. 
The relevant Feynman diagram is shown in Fig.~\ref{fig:zbb}(b). 
%
The one-loop effective Lagrangian that is generated by the Feynman diagram is
\bea
	{\cal L}_{\pi b\bar{b}} = \epsilon_b  \, \frac{2}{v} \bar{b}_L \gamma^\mu b_L \partial_\mu \pi^0,
	\label{eq:pibb}
\eea
where  
\bea
\epsilon_b &=& -\frac1{16\pi^2v^2}\Big[m_t^4c_L^4 C_0(m_t^2,m_t^2,0) 
+ m_T^4s_L^4 C_0(m_T^2,m_T^2,0)\nn\\
&+& 2m_t^2m_T^2c_L^2s_L^2 C_0(m_t^2,m_T^2,0)\Big],
\label{eq:epszbb}
\eea
Here $C_0(m_1^2, m_2^2, m_3^2)$ is the three-point Passarino-Veltman (PV) function~\cite{Passarino:1978jh}
in the zero external momentum limit, where $m_i$ are the 
masses of the particles in the triangle loop. 
In the limit of  the massless Goldstone boson, the three-point PV function
reduces to
\bea
C_0(m_1^2,m_2^2,0)=
\begin{cases} -\dfrac{1}{m_1^2-m_2^2}\ln\dfrac{m_1^2}{m_2^2} & \mbox{if }  m_1 \neq m_2 \\ 
-\dfrac{1}{m_1^2} & \mbox{if }  m_1 = m_2. \end{cases} 
\eea
In the decoupling limit, taking $s_L \to 0$ in Eq.~\ref{eq:epszbb}
 one recovers the leading $m_t$-dependent contribution in the SM:
\bea
	\epsilon_b^{\tbox{SM}} &=& \frac{m_t^2}{16\pi^2 v^2}.
\eea 
Based on the Ward identity in Ref.~\cite{Barbieri:1992dq, Fleischer:1994cb},  
we recognize the coefficient $\epsilon_b$ in Eq.~\ref{eq:pibb} 
 is proportional to the quantity we are interested in
\bea
	\delta g_{L}^{\tbox{NP}} = \epsilon_b - \epsilon_b^{\tbox{SM}}.
\eea
So we obtain the expression for the NP correction $\delta g_L^{\tbox{NP}}$:
\bea
\delta g_L^{\tbox{NP}} &=& \frac{m_t^2s_L^2}{16\pi^2v^2}\Big[ - (1 + c_L^2)\nn\\
&+& s_L^2 \frac{m_T^2}{m_t^2} + c_L^2 \frac{2 m_T^2}{m_T^2 - m_t^2} \ln\frac{m_T^2}{m_t^2}\Big].
\eea
Note that the terms inside the bracket are the same as in Eq.~\ref{eq:deltaTf}.  

\begin{figure}[t]
\begin{center}
\includegraphics[width=0.49\textwidth]{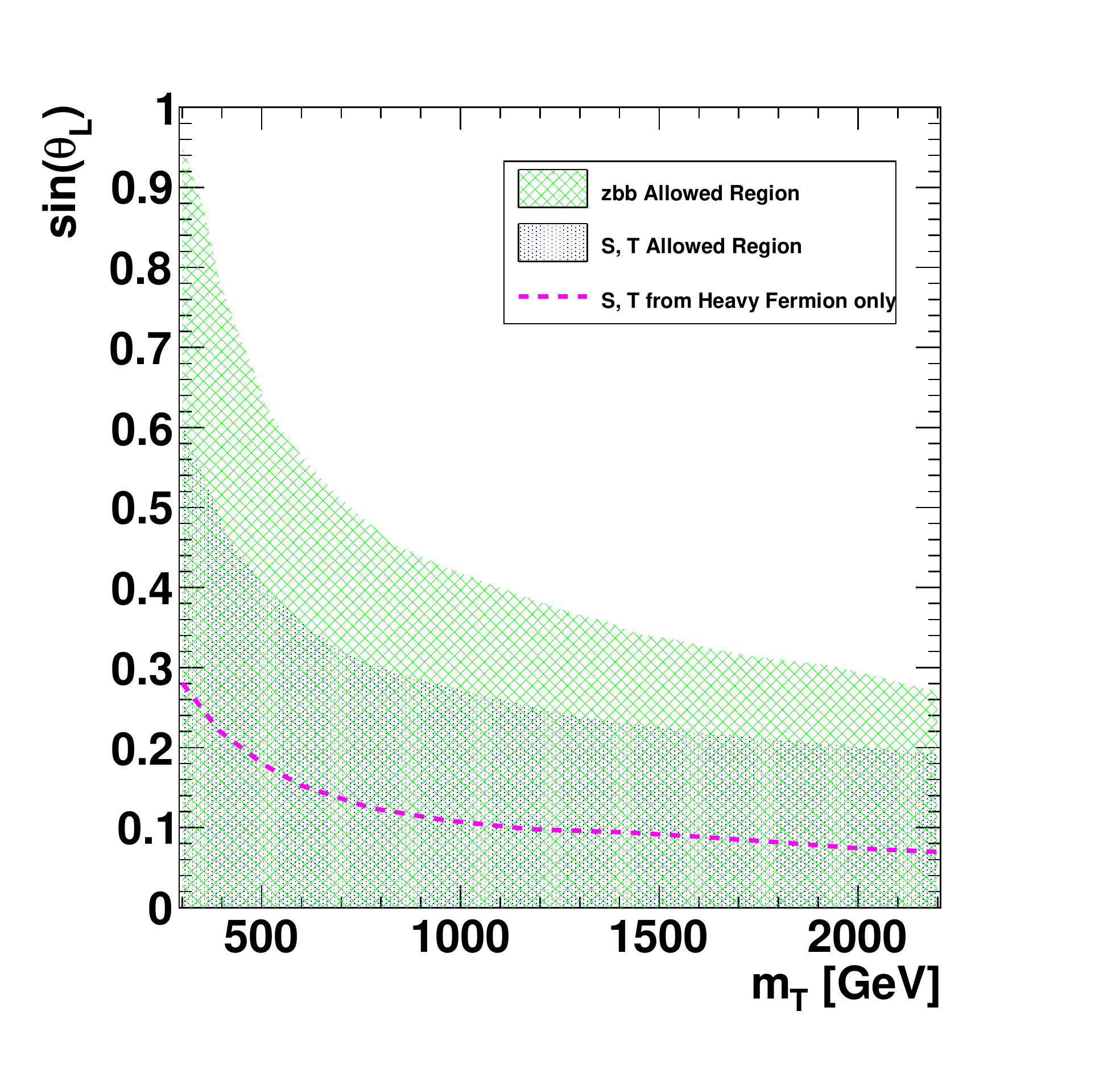} 
\caption{\small The allowed region from the oblique corrections $S, T$, and $Z\bar{b}_Lb_L$ at the 95\% confidence level. 
Below the pink dash line is the allowed region from the $S, T$ parameters in the pure vector-like fermion model. }
\label{fig:stzbb}
\end{center}
\end{figure}

Among all electroweak observables, 
three of them are related to the $Z\bar{b}b$ couplings:
${\cal A}_b$, $A_{FB}^{0,b}$,  and $R_b$.
It is known that the asymmetries ${\cal A}_b$ and $A_{FB}^{0,b}$ 
are mainly sensitive to $\delta g_{R}^{\tbox{NP}}$, while 
the $R_b$ mainly sets constraint on $\delta g_{L}^{\tbox{NP}}$. 
Due to the dominant corrections on the $\delta g_{L}^{\tbox{NP}}$, we 
will make use of the observable $R_b$ to constrain the parameter space. 
The shift in $R_b$ due to new physics is 
\bea
	\delta R_b  =  2 R_b (1 - R_b) \frac{g_L \delta g_L^{\tbox{NP}} + g_R \delta g_R^{\tbox{NP}}} {g_L^2 + g_R^2}.
\eea
The experimental value and SM theoretical value (including two loop corrections)~\cite{Beringer:1900zz} is
\bea
	R_b^{\tbox{exp}} = 0.21629 \pm 0.00066 \\
	R_b^{\tbox{th}} = 0.21575 \pm 0.00003 .
\eea
Following the Ref.~\cite{Ciuchini:2013pca},
$\delta g_{L}^{\tbox{NP}}$ is determined to be $\delta g_L = 0.0028 \pm 0.0014$.

The experimental constraints on the oblique parameters and the $Z\bar{b}b$ couplings
set limits on the parameter space in the model.
We scan over a large range of the parameter space, and 
obtain the allowed paramter region at the 95\% confidence level (CL). 
We find that all the parameter space on the $(m_S, s_\varphi)$ in the scalar sector are allowed.
This means the constraint from the electroweak precision data on the scalar sector is quite weak.
However, only part of the parameter space on the  $(m_T, s_L)$ in the top sector
is allowed, as is shown in Fig.~\ref{fig:stzbb}.
Since all the NP corrections are proportional to  $s_L^2$, there is a upper limit on the $s_L$.  
Fig.~\ref{fig:stzbb} shows the decoupling nature of the vector-like fermion:
as the fermion becomes heavier, there are less allowed region of the mixing angle.
The constraint from the non-universal correction to the $Zb\bar{b}$ coupling is weaker than 
the one from the universal oblique corrections.
The tightest constraint comes from the T parameter, since the  
vector-like fermion is in the singlet representation of the electroweak group, 
which contributes to the custodial symmetry breaking 
in the model at the loop level.
%
%
%
The dashed line in the Fig.~\ref{fig:stzbb} shows tighter constraints in the pure 
vector-like fermion model than in our model.
The relaxed constraint on T parameter in our model is due to 
the opposite correction from the boson loops with respect to the fermion contribution.
Therefore, the existence of the heavy scalar leads to 
larger allowed parameter space. 
%


\section{Higgs Coupling Measurements}
\label{sec:higgs}

Current data on the measurements of the coupling properties of the Higgs boson at the LHC
show that there is no significant deviation from the SM expectation. 
This put constraints on the NP models in which the Higgs couplings to the SM particles are modified. 
The deviations in the Higgs couplings can occur in two ways: 
new fermions or charged bosons contribute to 
the loop-induced $h\gamma\gamma$ and/or $hgg$ couplings;
new scalars mixed with the Higgs boson give rise to 
the deviation in the tree-level $hVV$ and/or $hf\bar{f}$ couplings.
In our model, the heavy scalar mixed with the Higgs boson
induces tree level correction to the couplings of the Higgs to the SM particles.
The vector-like fermion also contribute to the loop-induced $hgg$ coupling 
and $h\gamma\gamma$ coupling.
These effects will modify both the production cross section 
and the decay branching ratio of the Higgs boson. 
In the narrow-width approximation, the signal cross section 
can be decomposed in the following way for all channels
\bea
\left(\sigma\cdot\text{BR}\right)(\mathit{ii}\to h \to\mathit{ff}) 
\equiv \sigma_{\mathit{ii}}\cdot\text{BR}_{\mathit{ff}}
= \frac{\sigma_{\mathit{ii}}\cdot\Gamma_{\mathit{ff}}}{\Gamma_{h}} ,
\eea
where $\sigma_{\mathit{ii}}$ is the production cross section through the initial state $\mathit{ii}$, 
$\Gamma_{\mathit{ff}}$ the partial decay width into the final state $\mathit{ff}$ 
, $\Gamma_{h}$ the total width of the Higgs boson, and $\text{BR}_{\mathit{ff}}$ the branching ratio. 

\begin{figure}[t]
\begin{center}
\includegraphics[width=0.2\textwidth]{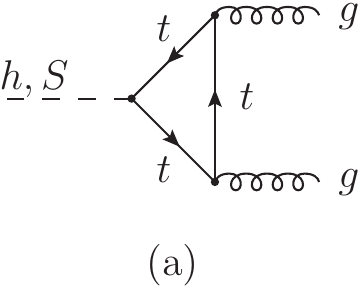} 
\includegraphics[width=0.2\textwidth]{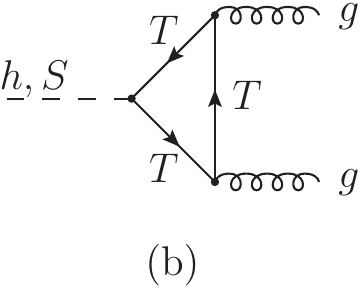} \\
\includegraphics[width=0.2\textwidth]{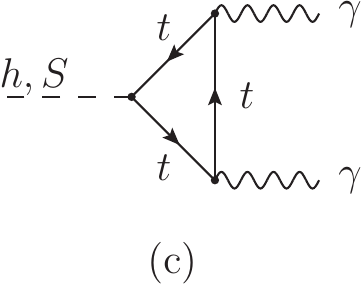} 
\includegraphics[width=0.2\textwidth]{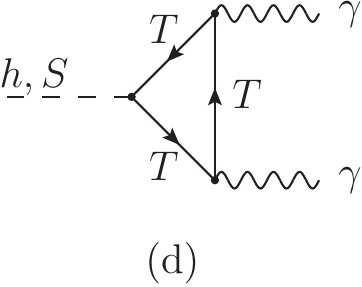} \\
\caption{\small The loop-induced $h\to gg$ Feynman diagrams due to the top quark (a) and the vector-like fermion (b), 
and  the loop-induced $h \to \gamma\gamma$  Feynman diagrams due to the top quark (c) and the vector-like fermion (d).}
\label{fig:higfeyn}
\end{center}
\end{figure}

Let us parametrize the deviations on the Higgs couplings 
in terms of the Higgs coupling scale factors $\kappa$, 
defined as $g^{\tbox{NP}}/g^{\tbox{SM}}$. 
The general effective Higgs couplings could be rewritten as
\bea
{\cal L}_{\textrm{Higgs}} &=& \kappa_{W} g_{\tbox{hWW}}^{\tbox{SM}} \,h W^{+\mu} W^{-}_\mu 
			+  \kappa_{Z} g_{\tbox{hZZ}}^{\tbox{SM}} \, h Z^\mu Z_\mu  \nn\\
			 &-& \kappa_t g_{htt}^{\tbox{SM}}\, h \bar{t}t 
			 - \kappa_b g_{hbb}^{\tbox{SM}}\, h \bar{b}b 
			 -\kappa_\tau g_{h\tau\tau}^{\tbox{SM}} \,h \bar{\tau}\tau \nn \\
			 &+& \kappa_g g_{h\gamma\gamma}^{\tbox{SM}}\, h G^{\mu\nu} G_{\mu\nu} 
			 + \kappa_\gamma g_{h gg}^{\tbox{SM}}\, h A^{\mu\nu} A_{\mu\nu},
\eea
where $g^{\textrm{SM}}$ are the SM Higgs couplings. 
At the tree level,  
$g_{hWW}^{\tbox{SM}} = \frac{2 m_W^2}{v}, g_{hZZ}^{\tbox{SM}} = \frac{m_Z^2}{v}$ 
and $g_{hff}^{\tbox{SM}} = \frac{m_f}{v}$.
On the other hand, the couplings $g_{h\gamma\gamma}^{\tbox{SM}}$ 
and $g_{h gg}^{\tbox{SM}}$ 
only receive loop corrections, and thus are suppressed by the loop factor. 
Up to one-loop level, the SM couplings to the photon and the gluon are
\bea
	g_{h gg}^{\tbox{SM}}          &=& \frac{g_s^2}{16\pi^2} \sum_{f} \frac{g_{hff}^{\tbox{SM}}}{m_f} A_{1/2}(\tau_f), \\
    g_{h\gamma\gamma}^{\tbox{SM}} &=& \frac{e^2}{16\pi^2}\left[\frac{g_{\tbox{hWW}}^{\tbox{SM}}}{m_W^2} A_{1}(\tau_W) 
	+ \sum_{f} 2 N_{c}^{f} Q_f^2 \frac{ g_{hff}^{\tbox{SM}} }{m_f} A_{1/2}(\tau_f)\right],\nn\\
\eea
where the sum over $f$ runs over $t, b, s, c$ quarks, and $\tau_i = \frac{4 m_i^2}{m_h^2}$.
Here the loop function $A_1(\tau)$ and $A_{1/2}(\tau)$ are
\bea
\label{eq:loop1}
A_{1/2}(\tau) &=& 2  \, \tau \left[ 1 + (1-\tau)f(\tau)\right] ,\\
A_1(\tau)&=& -2-3\tau\left[1+(2-\tau)f(\tau)\right].
\eea
with
\bea
f(x) &=& \begin{cases} \arcsin^2 [1/\sqrt{x}] \ , &\textrm{for $x \geq 1$}\ ,  \\ 
 -\frac{1}{4}\left[\ln\frac{1+\sqrt{1-x}}{1-\sqrt{1-x}} - i\pi\right]^2\ , &\textrm{for $x < 1$} .
 \end{cases}
\eea
In our model, due to mixing between the Higgs boson and the heavy scalar, all the 
tree-level Higgs couplings are modified as
\bea
	g_{h ff}^{\tbox{NP}} = c_\varphi g_{h ff}^{\tbox{SM}}, 
	\quad  g_{h VV}^{\tbox{NP}} = c_\varphi g_{h VV}^{\tbox{SM}}.
\eea
The loop-induced Higgs couplings to the photon and the gluon are also modified 
by the new contribution from the vector-like fermion loop, as shown in Fig.~\ref{fig:higfeyn}. 
So the Higgs couplings to the photon and the gluon are
\bea
	g_{h gg}^{\tbox{NP}}          &=&
	 \frac{g_s^2}{16\pi^2} \left( \sum_{f} \frac{g_{hff}}{m_f} A_{1/2}(\tau_f) + \frac{g_{hTT}}{m_T} A_{1/2}(\tau_T)\right), \\
    g_{h\gamma\gamma}^{\tbox{NP}} &=& \frac{e^2}{16\pi^2}\left[\frac{g_{\tbox{hWW}}}{m_W^2} A_{1}(\tau_W) 
	+ \sum_{f} 2 N_{c}^{f} Q_f^2 \frac{ g_{hff} }{m_f} A_{1/2}(\tau_f) \right. \nn\\
	&& \left.  +  \frac83 \frac{g_{hTT}}{m_T} A_{1/2}(\tau_T) \right],
	\label{eq:couphignew}
\eea
where $\tau_T = \frac{4 m_T^2}{m_h^2}$, and $g_{htt}, g_{hTT}$ couplings are given in Appendix A.
Note that in the above equations there is no scalar mass $m_S$ dependence.
From these couplings, we obtain the Higgs coupling scale factors $\kappa$ in the model:
\bea
	\kappa_{V} &=& \kappa_{f} = c_\varphi, \quad \kappa_{g} = \frac{ g_{h gg}^{\tbox{NP}} }{ g_{h gg}^{\tbox{SM}}  } ,
	\quad \kappa_{\gamma} = \frac{ g_{h \gamma\gamma}^{\tbox{NP}} }{ g_{h \gamma\gamma}^{\tbox{SM}}  } .	
\eea
The above parameters are not independent. 
$\kappa_{\gamma}$ can be expressed in terms of $\kappa_{g}$, 
because the loop contribution from the vector-like fermion to 
the $h \to \gamma\gamma$ and $h \to gg$ couplings are the same.
Therefore, there are only two independent parameters  ($\kappa_{V}, \kappa_{g}$) 
in our general parametrization of the Higgs couplings in the model.

\begin{figure}[t]
\begin{center}
\includegraphics[width=0.49\textwidth]{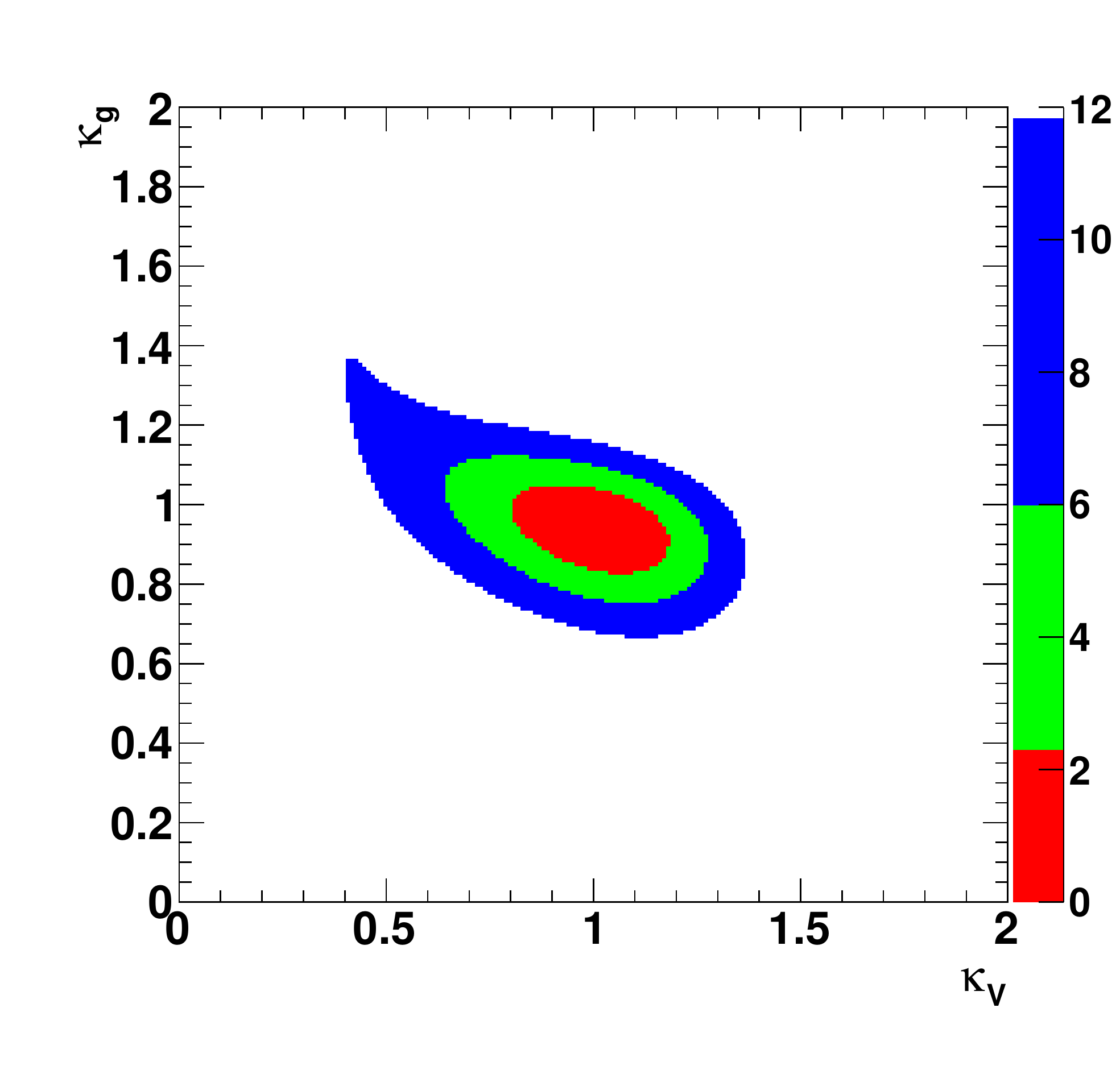} 
\caption{\small The allowed parameter region for the general parametrization $(\kappa_V, \kappa_{g})$ at the 68.27\% (red), 
95\% (green),  99.7\% (blue) confidence levels. The color pallette in the right side shows the 
value of the $\Delta \chi^2$ for the allowed parameter. }
\label{fig:higfit}
\end{center}
\end{figure}

\begin{figure}[t]
\begin{center}
\includegraphics[width=0.49\textwidth]{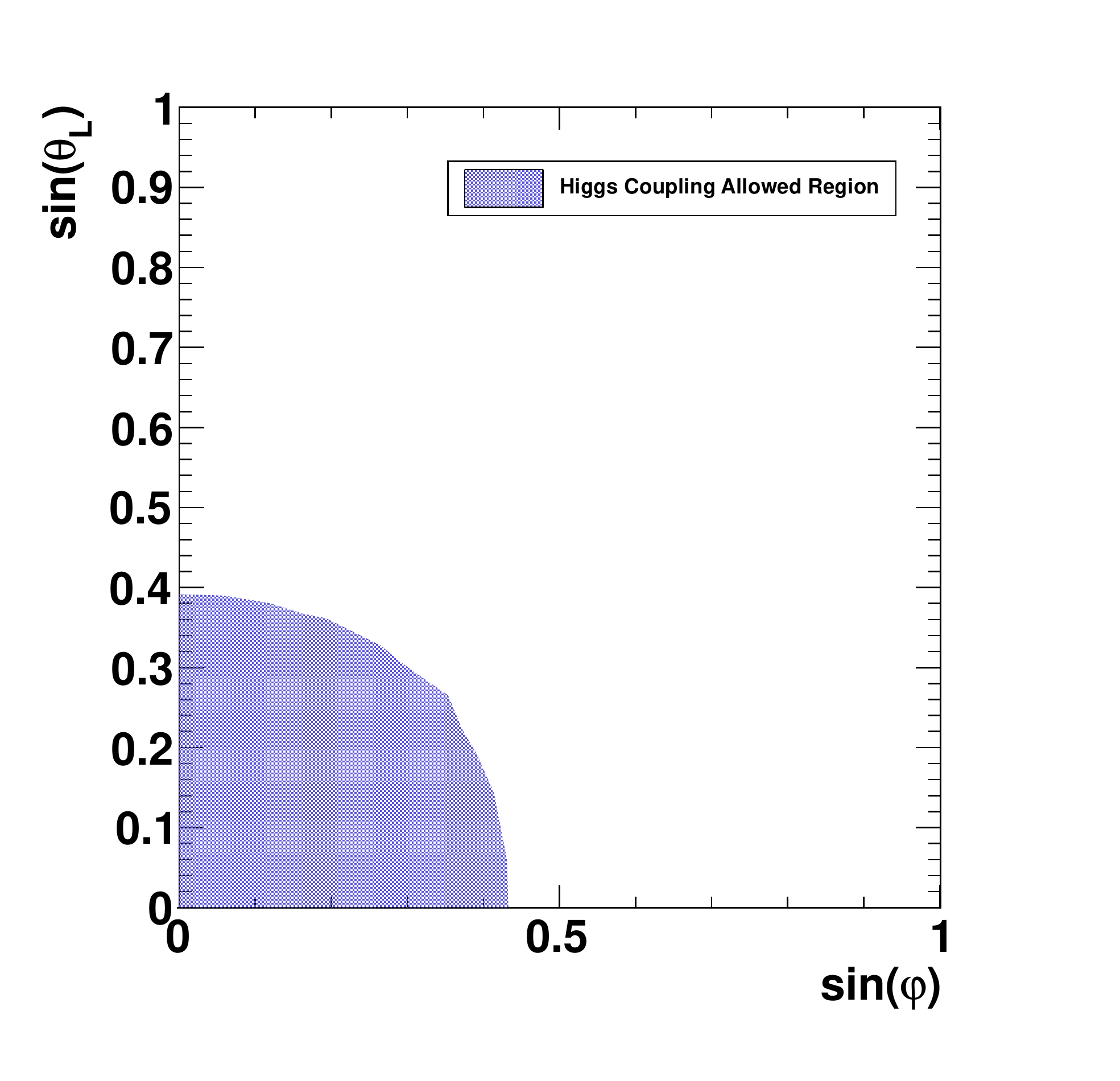} 
\caption{\small The allowed  parameter  region for the parameters ($s_\varphi, s_L$) in the model at the 95\% CL. }
\label{fig:higfit2}
\end{center}
\end{figure}

In the Higgs measurement at the LHC, 
the signal strength modifier~\cite{LHCHiggsCrossSectionWorkingGroup:2012nn} for each individual channel is defined
\bea
	R_{\mathit{ii}\to h \to\mathit{ff}} =  
	\frac{\sigma^{\tbox{NP}}_{\mathit{ii}}  \cdot \text{BR}^{\tbox{NP}}_{\mathit{ff}}  }
	{\sigma^{\tbox{SM}}_{\mathit{ii}} \cdot  \text{BR}^{\tbox{SM}}_{\mathit{ff}} }.
\eea
The main channels for the Higgs production are the gluon fusion process $gg\to h$, 
the vector-boson funsion (VBF) process $qq \to h qq$, and
the associated production $qq \to Vh$. 
To distinguish them, we denote the intial state $\mathit{ii}$ 
as (${\mathit{gg}}, \tbox{HV}, \tbox{VBF}$). 
So the production cross section in the parton level could be written as
\bea	
	\sigma^{\tbox{NP}}_{\mathit{gg}} = \kappa_{g}^2 \sigma^{\tbox{SM}}_{\mathit{gg}}, \quad 
	\sigma^{\tbox{NP}}_{\tbox{HV} }  =   \kappa_V^2 \sigma^{\tbox{SM}}_{\tbox{HV} }  , \quad
	\sigma^{\tbox{NP}}_{\tbox{VBF}}  =  \kappa_V^2  \sigma^{\tbox{SM}}_{\tbox{VBF}} .
\eea
In practice, it is hard to seperate contributions from different production channels.
So the production cross section in the hadron level should be
\bea
	\sigma_{pp} = \sum_i c_i \sigma^{\tbox{NP}}_{\mathit{ii}},
\eea
where $c_i$ is the relative contribution from the production channel $\mathit{ii}$, 
which depends on experimental cuts and detector effeciencies. 
At the current LHC running, the following decay channels: $ h\to WW/ZZ$, $ h\to \gamma\gamma$, 
and $ h\to bb/\tau\tau$ are observed.
Among these channels,
the measurements in the $ h\to bb/\tau\tau$ channels still have 
large uncertanties on the coupling measurements as well as poor mass resolution.
Therefore, the precise determination of the Higgs coupling mainly comes from 
the $ h\to WW/ZZ$, $ h\to \gamma\gamma$ channels. 
The decay branching ratio in the model is
\bea
	\text{BR}_{\mathit{\gamma\gamma}} &=& r_{\tbox{$\Gamma$}} \kappa_{\gamma}^2\text{BR}^{\tbox{SM}}_{\mathit{\gamma\gamma}}    ,\quad
	\text{BR}_{\mathit{VV}}           =   r_{\tbox{$\Gamma$}}  \kappa_{V}^2    \text{BR}^{\tbox{SM}}_{\mathit{VV}}               , \nn\\
	\text{BR}_{\mathit{ff}}           &=& r_{\tbox{$\Gamma$}} \kappa_{f}^2     \text{BR}^{\tbox{SM}}_{\mathit{ff}}               .
\eea
where $r_{\tbox{$\Gamma$}} = \frac{ \Gamma_h^{\tbox{SM}} } { \Gamma_h^{\tbox{NP}} }$ is 
the ratio of the total width in the SM against that in our model.
In terms of the scale factors in the model, the total decay width is parametrized as
\bea
	\Gamma_h^{\tbox{NP}}  \simeq (0.917 \kappa_V^2 + 0.003 \kappa_\gamma^2 + 0.08 \kappa_g^2 )\Gamma_h^{\tbox{SM}},
\eea
where the coefficients come from the numerical values of the SM contributions.

Given the total production cross section and the decay branching ratios 
in terms of the two independent parameters  ($\kappa_{V}, \kappa_{g}$),
we are ready to compare our theoretical predictions with the experimental data.
We will use all the availiable data from the Table 10 and 11 in the Ref.~\cite{Bechtle:2013xfa}, 
which collect both the ATLAS and CMS experimental results with full integrated luminosity at the 7 TeV and 8 TeV.
Then we perform a global fit based on the $\chi^2$ analysis:
\bea
	\chi^2 = \sum_i \left( \frac{ R^{\tbox{exp}}_i - R^{\tbox{th}}_i }
	{\sigma^{\tbox{exp}}_i}\right)^2,
\eea
where we sum over all the available channels in the measurements. 
Here $R^{\tbox{exp}}$ is the Higgs signal modifier obtained from the experimental data, 
and $\sigma^{\tbox{exp}}$ is its experimental error.
Fig.~\ref{fig:higfit} shows the allowed parameter region for the scale factors $(\kappa_V, \kappa_{g})$ at the 68.27\%, 
95\%,  99.7\% CLs, respectively. 
The best fit of the scale factors ($\kappa_{V}, \kappa_{g}$) is found to be
\bea
	\kappa_{V} = 1.0096 \pm 0.297, \quad  \kappa_{g} =  0.941 \pm 0.176, \nn\\
	 \textrm{with } \chi^2/\textrm{ndf} = 61.022/54.
	 \label{eq:bestfit}
\eea
From Eq.~\ref{eq:bestfit} and Fig.~\ref{fig:higfit},
we find that the central value of the best fit is close to the SM value,
and the 95\% CL contour shows a moderate accuracy. 
In Fig.~\ref{fig:higfit}, there is a small tail in the  99.7\% CL contour.
This tail tells us the parameter space with larger $\kappa_g$ and smaller $\kappa_V$ 
is still allowed by the current data.
In the $h \to \gamma\gamma$ channel, due to enhanced total cross section,
$\kappa_g$ could be larger.
While in the $h \to VV$ channel, since the total cross section $\sigma_{gg \to h \to VV} \sim \kappa_g^2 \kappa_V^2$,
a larger $\kappa_g$ implies a smaller $\kappa_V$.
%
%
%
We could convert our constraints on the scale factors ($\kappa_{V}, \kappa_{g}$) into a limit on the model parameters, 
as shown in Fig.~\ref{fig:higfit2}.
From the Fig.~\ref{fig:higfit2}, we read that the large mixing angles ($s_\varphi, s_L$) are not allowed. 
The constraint on $s_\varphi$ is expected since all the NP corrections in the total cross section are proportional to $s_\varphi^2$. 
%
The constraint on $s_L$ mainly comes from the gluon fusion production cross section and the $h \to \gamma\gamma$ decay branching ratio.
The constraints on the parameters ($u$, $m_T$) are quite weak, and there is no constraint on $m_S$ at all.
For the scale $u$, it only appears in the combination $\frac{s_\varphi}{u}$, 
inside the $gg \to h$ and $h \to \gamma\gamma$ loops.
Since $s_\varphi$ can not be large, there is no constraint for a TeV scale $u$.
%
Due to the saturated behavior of the function $A_{1/2}(\tau_T)$ for a heavy $m_T$,
the constraint on the $m_T$ is weak.


\section{Hadron Collider Searches}
\label{sec:collider}

There are many direct searches on the vector-like quarks which couple predominantly
to the third-generation quarks at the Tevatron and the LHC.
At the LHC, the vector-like quark could be produced in pair through QCD production $pp\to T\bar{T}$, 
or be singly produced via electroweak process $pp \to T\bar{b}$. 
For a light vector-like fermion, the pair production cross section is larger than the one in the single production,
while  for a heavy vector-like fermion the single production is more efficient. 
The main decay channels of the heavy vector-like fermion are
\bea
	T \to t Z, \quad T \to b W, \quad T \to t h,
\eea
and $T \to t S$ only if the scalar is much lighter than the vector-like fermion.
In the model, the tree-level partial decay widths are given by
\bea
	\Gamma_\mathit{T \to bW} &=& \frac{s_L^2 m_T^3}{32\pi v^2} 
	\left(1 + {\cal O}(\frac{m_W^2}{m_T^2}) \right), \\
	\Gamma_\mathit{T \to tZ} &=& \frac{s_L^2 m_T^3}{64\pi v^2} 
	\left((1 - \frac{m_t^2}{m_T^2})^3 + {\cal O}(\frac{m_Z^2}{m_T^2}) \right), \\
	\Gamma_\mathit{T \to th} &=& \frac{s_L^2 c_L^2 c_\varphi^2 m_T^3}{64\pi v^2}
	\left( 1 + \frac{5 m_t^2}{m_T^2}  + {\cal O}(\frac{m_h^2}{m_T^2},\frac{m_t^4}{m_T^4}, 
	\frac{v^2}{u^2}) \right).\nn\\
\eea
Taking the limit $m_T \gg m_t, m_h$, the partial decay widths have the following pattern:
\bea
	\Gamma_\mathit{T \to bW} : \Gamma_\mathit{T \to tZ} : \Gamma_\mathit{T \to th} \simeq 2:1:1.
\eea
This can be understood in the following way.
Using the Goldstone equivalent theorem~\cite{Yao:1988aj, Bagger:1989fc, He:1992nga}, the partial decay widths 
can be  estimated by calculating the corresponding Goldstone boson final state instead:
\bea
\Gamma_\mathit{T \to bW} \simeq \Gamma_\mathit{T \to b\pi^\pm} , 
\quad \Gamma_\mathit{T \to tZ} \simeq \Gamma_\mathit{T \to t\pi^0}.\nn
\eea 
Due to the custodial symmetry in the scalar sector, 
the partial decay widths have 
\bea
	\Gamma_\mathit{T \to b\pi^\pm} : \Gamma_\mathit{T \to t\pi^0} : \Gamma_\mathit{T \to th} \simeq 2:1:1.
\eea
In a recent CMS analysis~\cite{Chatrchyan:2013uxa}, using the 8 TeV data collected 
up to integrated luminosity of $19.5 $ fb$^{-1}$,
the up-to-dated lower limits on the mass of the heavy fermion are set to be around $687 - 782$ GeV 
depending on different patterns 
of the vector-like quark decay branching ratios. 
%
%
Setting the pattern of the branching ratio as $2:1:1$, 
we could put a limit on the vector-like fermion mass: $696$ GeV. 

\begin{figure}[t]
\begin{center}
\includegraphics[width=0.47\textwidth]{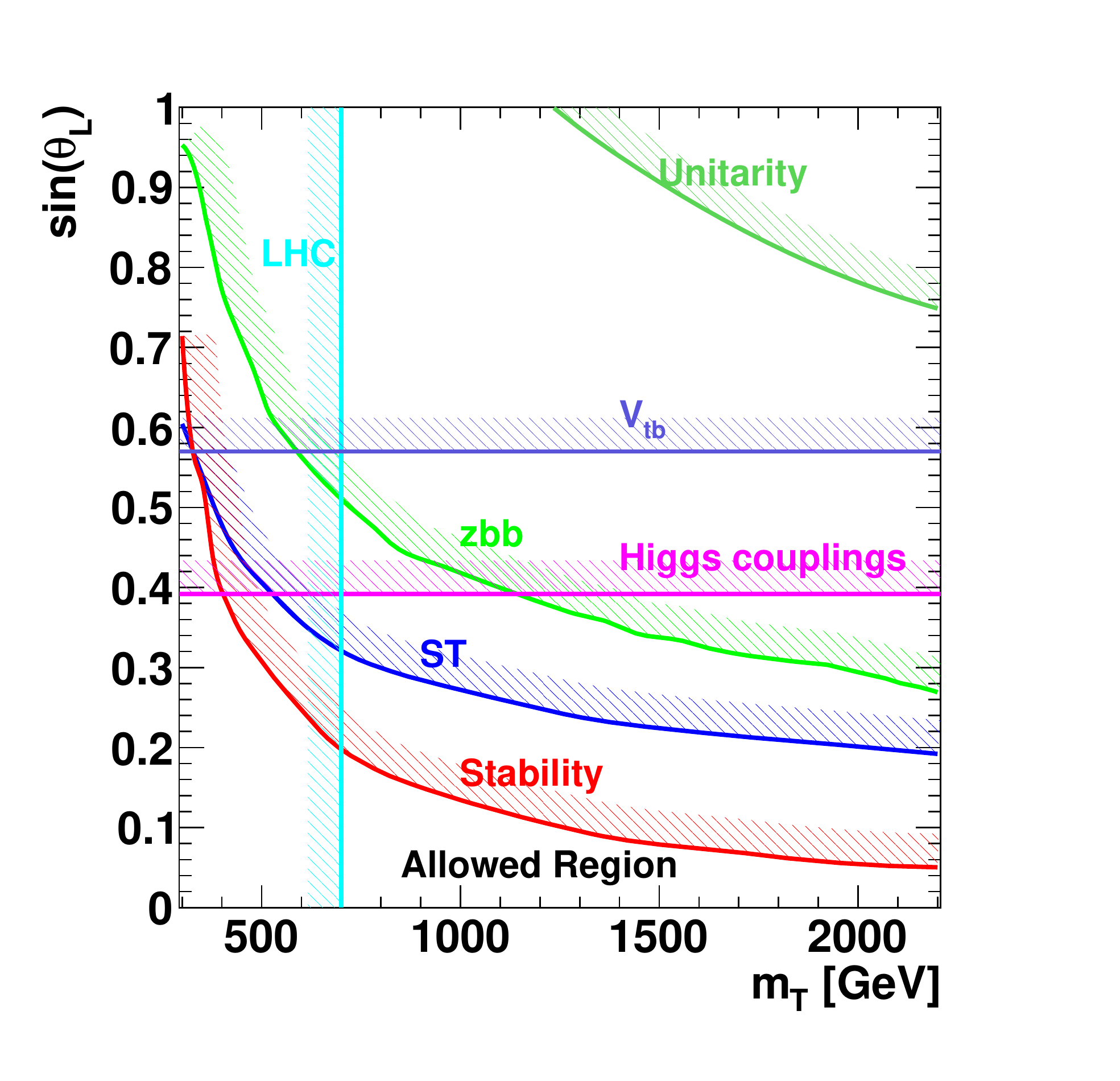} 
\caption{\small Exclusion plot on the parameters $m_T - \sin\theta_L$ with all current constraints included. 
The exclusion zones are on the shadow side of each line. The allowed region is shown as the zone delimited by the tightest constraints from the Stability and the LHC. }
\label{fig:tot_mf}
\end{center}
\end{figure}

\begin{figure}[t]
\begin{center}
\includegraphics[width=0.47\textwidth]{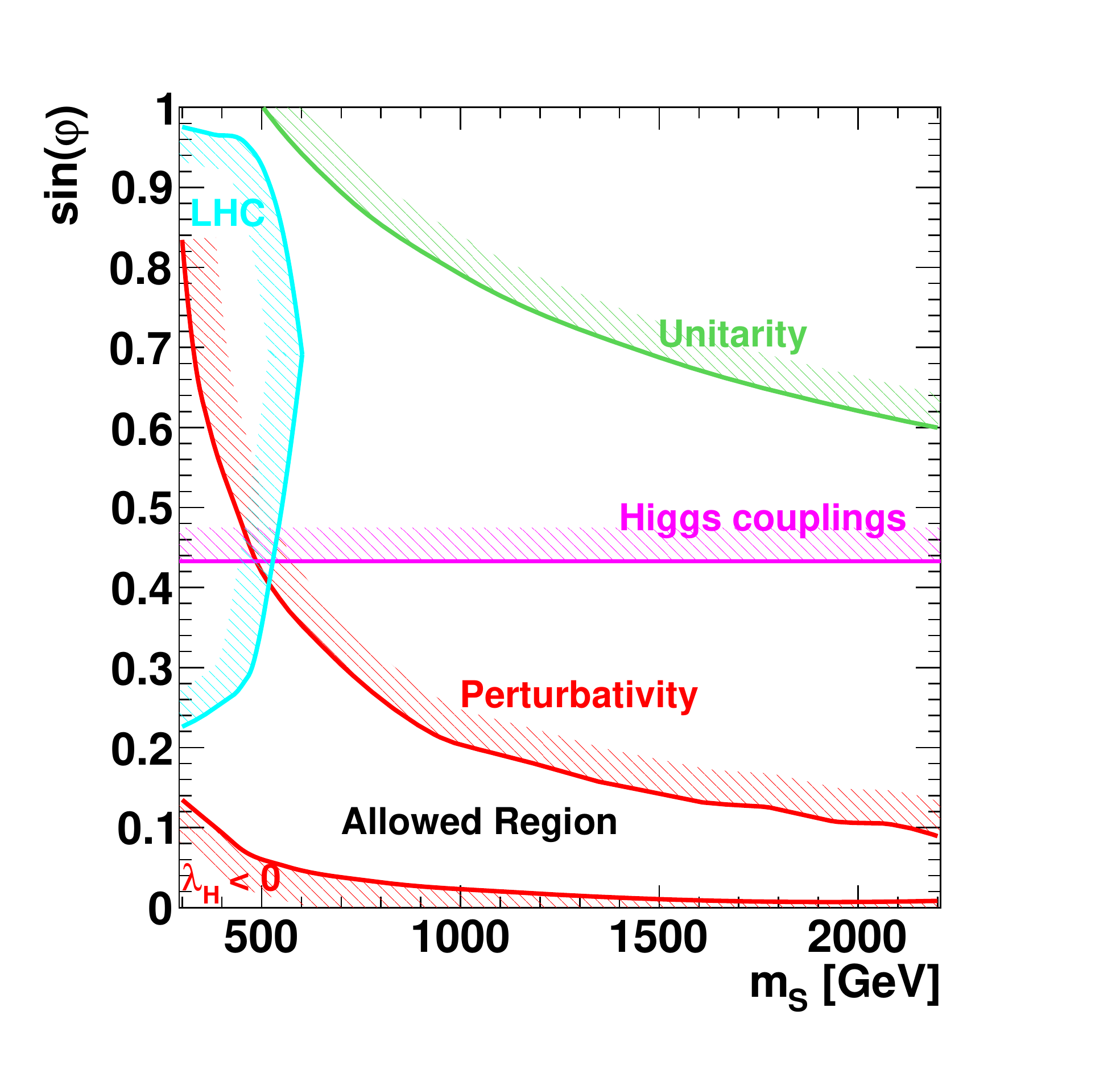} 
\caption{\small Exclusion plot on the parameters $m_S - \sin\varphi$  with all current constraints included. 
The exclusion zones are on the shadow side of each line. The allowed region is shown  as the zone  delimited by the tightest constraints from the Perturbativity, the Stability ($\lambda_H < 0$), and the LHC. }
\label{fig:tot_ms}
\end{center}
\end{figure}

In the model, the heavy scalar is CP-even, the same as the SM Higgs boson. 
The search limits on the high mass Higgs boson at the Tevatron and the LHC
could be used to set constraints on the mass and couplings of the heavy scalar.
%
%
%
The production mechanism is similar to the Higgs boson, 
dominanted by the gluon fusion with the production cross section $\sigma_{gg \to S}$. 
The decay channels of the heavy scalar are 
\bea
	S \to WW, \quad S \to ZZ, \quad S \to hh, \quad S \to t\bar{t},
\eea
and  $S \to t T$ only if the vector-like fermion is much lighter than the scalar.
Other decay channels, such as $S \to \gamma \gamma/gg$, $S \to f\bar{f}$, 
where $f$ is the fermion other than the top quark, are negligible. 
The explicit formulae of the partial decay widths are listed in the Appendix D.
Similar to the vector-like fermion case,
in the limit $m_S \gg m_h, m_t$, one could estimate the partial widths by calculating 
the decays to the corresponding Goldstone bosons
instead: 
\bea
	\Gamma_\mathit{S \to WW} &\simeq& \Gamma_\mathit{S \to \pi^+\pi^-} 
	= \frac{s_\varphi^2 m_S^3}{32 \pi v^2} \left(1 + {\mathcal O}(\frac{m_Z^2}{m_S^2})\right),\\
	\Gamma_\mathit{S \to ZZ} &\simeq& \Gamma_\mathit{S \to \pi^0 \pi^0} 
	= \frac{s_\varphi^2 m_S^3}{64 \pi v^2} \left(1 + {\mathcal O}(\frac{m_Z^2}{m_S^2})\right),\\
	\Gamma_\mathit{S \to hh} &=& \frac{s_\varphi^2 c_\varphi^4 m_S^3}{64 \pi v^2}  
	\left(1 + {\mathcal O}(\frac{m_h^2}{m_S^2},\frac{v^2}{u^2})\right).
\eea
Similar decay pattern holds here: 
$\Gamma_\mathit{S \to WW}: \Gamma_\mathit{S \to ZZ}: \Gamma_\mathit{S \to hh} \simeq 2:1:1$.
%
%
In an up-to-dated analysis from the CMS~\cite{Chatrchyan:2013yoa}, 
the searches in $S \to WW$ and $S \to ZZ$ decay channels are studied in the mass range between $145$ GeV and $1000$ GeV.
If the high mass Higgs boson has the same coupling as the SM, 
the mass range between $145$ GeV and $710$ GeV are excluded at the 95\% CL. 
We convert this constraint into the limit on the heavy scalar in the model.
After calculating the production cross section and the decay branching ratios of the heavy scalar,
we perform a scan over the whole range of the parameter space, which is shown in Fig.~\ref{fig:tot_ms}.
It is shown that a range of the parameter space with light scalar mass and moderate mixing angle
is ruled out.


\section{Conclusions}
\label{sec:conclusion}

We investigated a vector-like fermion coupled to a new singlet scalar and the third generation quarks. 
The singlet scalar extended the Higgs sector, through the mixing with the Higgs boson. 
In our setup, the mass of the vector-like fermion is purely generated from symmetry breaking of the singlet scalar.
%
We carefully examined the electroweak vacuum stability and scalar perturbativity via the 
renormalization group evolution of the Higgs quartic coupling, and the scalar quartic couplings.
The matching condition when integrating out heavy particles,
and the relation between the running and physical parameters, were considered.
Although the vector-like fermion 
provides negative contributions to the running of the Higgs quartic coupling, the new scalar  
contributes positively.
In the matching of the renormalization group, 
the Higgs quartic coupling obtains a positive threshod shift 
at the scale of the scalar mass.
Taking the above two effects into account,
it is likely that the Higgs quartic coupling could stay positive up to the Planck scale.  
We performed a  scan over  the parameter space,
and found that a large range of the parameter space is allowed.
In this model, we also examined the constraints from 
the precision electroweak observables, Higgs coupling precision measurements, 
and the LHC direct searches. 
In Fig.~\ref{fig:tot_mf}, and Fig.~\ref{fig:tot_ms}, 
we summarized current constraints on the parameter space of the top sector $(m_T, \sin\theta_L)$ 
and the scalar sector $(m_S, \sin\varphi)$. 
We also included the constraint from the $V_{tb}$ measurement~\cite{Beringer:1900zz}.
It is interesting to see that the tightest constraints always come from the Higgs vacuum stability 
and perturbativity in the $(m_T, \sin\theta_L)$ and $(m_S, \sin\varphi)$ spaces.
However, the constraints from the perturbative unitarity are very weak.
Concerning  the $(m_T, \sin\theta_L)$ parameter space,
the oblique parameters $S, T$ put a tight constraint on the parameter space,
while the constraints from the $Zb_L\bar{b}_L$ coupling and the Higgs coupling measurements are weaker.
The direct LHC searches set the lower limits on the mass of the vector-like fermion at around $700$ GeV.
On the other hand, there is no constraint  
on the $(m_S, \sin\varphi)$ parameter space 
from the oblique parameters $S, T$ and $Zb_L\bar{b}_L$ coupling measurements.
The Higgs coupling measurements give rise to an upper bound on the mixing angle $s_\varphi$.
Regarding to the LHC direct searches, 
only a small range of the parameter space with moderate mixing angle and light mass is ruled out.
As shown in Fig.~\ref{fig:tot_mf} and Fig.~\ref{fig:tot_ms}, a large region of the parameter space is still unexplored. 
%
%
%

\section*{Acknowledgements}

We would like to thank Can Kilic, Duane Dicus, Jacques Distler, and Willy Fischler for helpful discussion and valuable comments. 
The research of JHY was supported by the National Science Foundation under Grant Numbers PHY-1315983 and PHY-1316033.


%
%
%
%
%


\appendix


\begin{widetext}

\section{Relavent Electroweak Lagrangian in the Model}
\label{sec:appen1}

Let us summarize the relevant Lagrangian as follows. 
The interactions of the SM quarks $(t,b)$ reads
\bea
	\mathcal{L}_{W} &=& - \frac{g_2}{\sqrt{2}} \overline{t} \gamma^\mu \left(c_L P_L \right)b W^+_\mu + h.c. \,,\\
	\mathcal{L}_{Z} &=& - \frac{g_2}{c_W} \overline{t} \gamma^\mu \left(c_L^2 T^3_u P_L - Q s^2_W\right) t Z_\mu\,,\\
	\mathcal{L}_{H} &=& - \frac{\sqrt{2}\pi^+}{v} \overline{t} \left(m_bc_L P_R - m_tc_L P_L\right) b    + h.c.
					  - \left(\frac{m_tc_L^2}{v}(\phi - i \pi^0) + \frac{m_t s_L^2}{u}\chi \right) \overline{t} t\,,
\eea
where $\phi = c_\varphi h + s_\varphi S$, and $\chi = - s_\varphi h + c_\varphi S$.
The interactions with the gluon and the photon that are the same as in the SM.
The interactions of the heavy quark $T$ are
\bea
	\mathcal{L}_{Z} &=& - \frac{g_2}{c_W} \overline{T} \gamma^\mu (s_L^2 T^3_u P_L - Q s^2_W) T Z_\mu\,,\\
	\mathcal{L}_{H} &=& - \left( \frac{m_Ts_L^2}{v}(\phi - i \pi^0)+ \frac{m_T c_L^2}{u}\chi\right)\overline{T}T \,.
\eea
Finally, the terms involving in a $T$ and a $(t,b)$ are
\bea
	\mathcal{L}_{W} &=& - \frac{g_2}{\sqrt{2}} \overline{T} \gamma^\mu\left( s_L P_L \right) b W^+_\mu + h.c. \,,\\
	\mathcal{L}_{Z} &=& - \frac{g_2}{c_W} \overline{t} \gamma^\mu (s_L c_L T^3_u P_L) T Z_\mu + h.c.\,,\\
	\mathcal{L}_{H} &=& - \frac{\sqrt{2}\pi^+}{v}\overline{T}  (m_bs_L P_R - m_Ts_L P_L) b  
					 - \left(\frac{\phi - i \pi^0}{v}-\frac{\chi}{u}\right)\overline{t} \left(m_ts_L c_LP_L  + m_Ts_L c_LP_R  \right) T  + h.c.\,.
\eea

\section{Renormalization Group Equations}

In this section, we listed the one-loop RGEs in different effective field theories.

\subsection{Renormalization Group Equations in our Model}

At the scale $\mu > M_S, M_T$, both the heavy scalar and the vector-like fermion are involved in the RGE running.
The gauge coupling RGEs are 
\bea
	\frac{dg_1^2}{d\ln\mu^2} &=& \frac{g_1^4}{(4\pi)^2} \bigg[\frac{41 }{10} +\dfrac{16}{15}\bigg], \\
	\frac{dg_2^2}{d\ln\mu^2} &=& \frac{g_2^4}{(4\pi)^2} \bigg[-\frac{19}{6} \bigg], \\
	\frac{dg_3^2}{d\ln\mu^2} &=& \frac{g_3^4}{(4\pi)^2} \bigg[-7 +\dfrac{2}{3} \bigg],
\eea
where $g_1^2 = 5 g_Y^2/3$ is the hypercharge gauge coupling in GUT normalisation.
The Yukawa coupling RGEs are
\bea
\frac{dy_t^2}{d\ln\mu^2}&=& \frac{y_t^2}{(4\pi)^2} \bigg[ \frac{9 y_T^2}{2}+ \frac{9 y_t^2}{2} +\frac{3 y_b^2}{2}+y_{\tau }^2
-\frac{17 g_1^2}{20}-\frac{9 g_2^2}{4}-8 g_3^2 \bigg], \\
\frac{dy_b^2}{d\ln\mu^2}&=& \frac{y_S^2}{(4\pi)^2} \bigg[ \frac{3 y_t^2}{2}+\frac{9 y_b^2}{2}+ \frac{3 y_T^2}{2}+y_{\tau }^2
-\frac{g_1^2}{4}-\frac{9 g_2^2}{4} -8 g_3^2\bigg], \\
\frac{dy_\tau^2}{d\ln\mu^2}&=& \frac{y_\tau^2}{(4\pi)^2} \bigg[  3 y_t^2+3 y_b^2+3 y_T^2+\frac{ 5 y_{\tau }^2}{2}
-\frac{9 g_1^2}{4}-\frac{9 g_2^2}{4}\bigg], \\
\frac{dy_T^2}{d\ln\mu^2}&=&\frac{y_T^2}{(4\pi)^2} \bigg[ \frac{9}{2} y_T^2 + \frac92 y_t^2  +\frac32 y_b^2 +y_{\tau }^2  +\frac14 y_M^2  
-\frac{17}{20}g_1^2 - \frac{9}{4}g_2^2 - 8 g_3^2) \bigg], \\
\frac{dy_M^2}{d\ln\mu^2}&=&\frac{y_M^2}{(4\pi)^2} \bigg[ y_T^2  + \frac{9}{2} y_M^2
-\frac{8}{5}g_1^2 - 8 g_3^2  \bigg].
\eea
The RGEs in the Higgs sector are
\bea
\frac{d\lambda_H}{d\ln\mu^2}&=& \frac{1}{(4\pi)^2} \bigg[ \lambda_H \bigg( 12 \lambda_H  +6 y_t^2 + 6 y_b^2 + 2 y_\tau^2 +6 y_T^2 -\frac{9 g_1^2}{10}-\frac{9 g_2^2}{2} \bigg) \nn\\
&&  +  \bigg( \frac14 \lambda_{SH}^2 -3 y_t^4 - 3 y_b^4 - y_{\tau}^4  -3 y_T^4   - 6y_t^2 y_T^2  + \frac{27 g_1^4}{400}+\frac{9 g_2^4}{16}+\frac{9 g_2^2 g_1^2}{40}\bigg)  \bigg],\\
\frac{d\lambda_{SH}}{d\ln\mu^2}&=& \frac{1}{(4\pi)^2} \bigg[  \lambda_{SH} \bigg( 2\lambda_{SH}  + 6 \lambda_H +  3 \lambda_S 
 + 3 y_t^2 + 3 y_b^2 + y_\tau^2  +3 y_T^2 +  3 y_M^2 -\frac{9 g_1^2}{20}-\frac{9 g_2^2}{4}\bigg)   - 6  y_T^2y_M^2 \bigg], \\
\frac{d\lambda_S}{d\ln\mu^2}&=& \frac{1}{(4\pi)^2} \bigg[ 9 \lambda_S^2 + 6 y_M^2 \lambda_S + \lambda_{SH}^2 - 3 y_M^4  \bigg].
\eea

\subsection{The Standard Model + Vector-like Fermion Singlet}

At the scale $\mu <  M_S$, and $\mu > M_T$, only the vector-like fermion is involved in the RGE running.
The gauge coupling RGEs are
\bea
	\frac{dg_1^2}{d\ln\mu^2} &=& \frac{g_1^4}{(4\pi)^2} \bigg[\frac{41 }{10} +\dfrac{16}{15}\bigg], \\
	\frac{dg_2^2}{d\ln\mu^2} &=& \frac{g_2^4}{(4\pi)^2} \bigg[-\frac{19}{6}  + 0 \bigg], \\
	\frac{dg_3^2}{d\ln\mu^2} &=& \frac{g_3^4}{(4\pi)^2} \bigg[-7  + \dfrac{2}{3} \bigg].
\eea
The Yukawa coupling RGEs are
\bea
\frac{dy_t^2}{d\ln\mu^2}&=& \frac{y_t^2}{(4\pi)^2} \bigg[  \frac{9y_T^2}2  + \frac{9y_t^2}{2} +\frac{3 y_b^2}{2}+y_{\tau }^2
-\frac{17 g_1^2}{20}-\frac{9 g_2^2}{4}-8 g_3^2 \bigg], \\
\frac{dy_b^2}{d\ln\mu^2}&=& \frac{y_b^2}{(4\pi)^2} \bigg[ \frac{3 y_t^2}{2} + \frac32 y_T^2 +\frac{9 y_b^2}{2}+y_{\tau }^2
-\frac{g_1^2}{4}-\frac{9 g_2^2}{4} -8 g_3^2\bigg], \\
\frac{dy_\tau^2}{d\ln\mu^2}&=& \frac{y_\tau^2}{(4\pi)^2} \bigg[  3 y_t^2 + 3 y_T^2  +3 y_b^2+\frac{ 5 y_{\tau }^2}{2}
-\frac{9 g_1^2}{4}-\frac{9 g_2^2}{4}\bigg], \\
\frac{dy_T^2}{d\ln\mu^2}&=&\frac{y_T^2}{(4\pi)^2} \bigg[  \frac{9}{2} y_T^2 + \frac{9}{2} y_t^2 +\frac32 y_b^2 + y_{\tau}^2 
-\frac{17}{20}g_1^2 - \frac{9}{4}g_2^2 - 8 g_3^2) \bigg],
\eea
and the RGE in the Higgs sector is
\bea
&&\frac{d\lambda}{d\ln\mu^2} = \frac{\lambda}{(4\pi)^2} \bigg[ 12 \lambda +  6 y_T^2 +6 y_t^2+6 y_b^2+2 y_{\tau }^2
-\frac{9 g_1^2}{10}-\frac{9 g_2^2}{2}\bigg]  \nn\\
&&+ \frac{1}{(4\pi)^2} \bigg[ - 3 y_T^4 -   6 y_T^2 y_t^2 -3 y_t^4-3 y_b^4-y_{\tau }^4  +\frac{27 g_1^4}{400}+\frac{9 g_2^4}{16}+\frac{9 g_1^2 g_2^2}{40} \bigg].
\eea

\subsection{The Standard Model + Real Scalar Singlet}

At the scale $\mu >  M_S$, and $\mu < M_T$, only the vector-like fermion is involved in the RGE running.
The gauge coupling RGEs  are
\bea
	\frac{dg_1^2}{d\ln\mu^2} &=& \frac{g_1^4}{(4\pi)^2} \bigg[\frac{41 }{10}\bigg], \\
	\frac{dg_2^2}{d\ln\mu^2} &=& \frac{g_2^4}{(4\pi)^2} \bigg[-\frac{19}{6}\bigg], \\
	\frac{dg_3^2}{d\ln\mu^2} &=& \frac{g_3^4}{(4\pi)^2} \bigg[-7\bigg].
\eea
The Yukawa coupling RGEs are
\bea
\frac{dy_t^2}{d\ln\mu^2}&=& \frac{y_t^2}{(4\pi)^2} \bigg[ \frac{9 y_t^2}{2}+\frac{3 y_b^2}{2}+y_{\tau }^2
-\frac{17 g_1^2}{20}-\frac{9 g_2^2}{4}-8 g_3^2 \bigg], \\
\frac{dy_b^2}{d\ln\mu^2}&=& \frac{y_b^2}{(4\pi)^2} \bigg[ \frac{3 y_t^2}{2}+\frac{9 y_b^2}{2}+y_{\tau }^2
-\frac{g_1^2}{4}-\frac{9 g_2^2}{4} -8 g_3^2\bigg], \\
\frac{dy_\tau^2}{d\ln\mu^2}&=& \frac{y_\tau^2}{(4\pi)^2} \bigg[ 3 y_t^2+3 y_b^2+\frac{ 5 y_{\tau }^2}{2}
-\frac{9 g_1^2}{4}-\frac{9 g_2^2}{4}\bigg],
\eea
and the RGEs in the Higgs sector are
\bea
\frac{d\lambda_H}{d\ln\mu^2}&=& \frac{\lambda_H}{(4\pi)^2} \bigg[ 12 \lambda_H  +6 y_t^2+6 y_b^2+2 y_{\tau }^2
-\frac{9 g_1^2}{10}-\frac{9 g_2^2}{2}\bigg]  \nn\\
&+& \frac{1}{(4\pi)^2} \bigg[ \frac14 \lambda_{SH}^2 -3 y_t^4-3 y_b^4-y_{\tau }^4 +\frac{9 g_2^4}{16}+\frac{27 g_1^4}{400}+\frac{9 g_2^2 g_1^2}{40} \bigg], \\
\frac{d\lambda_{SH}}{d\ln\mu^2}&=& \frac{\lambda_{SH}}{(4\pi)^2} \bigg[ 2\lambda_{SH} + 6 \lambda_H + 3 \lambda_S + 3 y_t^2 + 3 y_b^2 +   y_\tau^2-\frac{9 g_1^2}{20}-\frac{9 g_2^2}{4}\bigg], \\
\frac{d\lambda_S}{d\ln\mu^2}&=& \frac{1}{(4\pi)^2} \bigg[ 9 \lambda_S^2 + \lambda_{SH}^2\bigg].
\eea

\section{Calculation of the Oblique Parameters $S, T$}

In this section we present the computation of the $S, T$ parameters 
with Passarino-Veltman functions~\cite{Passarino:1978jh}. 

\subsection{General Formulae For Gauge Boson Self-Energy}

We list the general formulae for the gauge boson self-energy functions $\Pi_{ij}$, where
$i, j$ denote the gauge boson species. 
%
In the formulae, only the one-point PV function $A_0(m^2)$, and the two-point PV functions 
$B_0(p^2, m_1^2, m_2^2), B_{00}(p^2, m_1^2, m_2^2)$ are involved.
In the calculation of the oblique parameters, all the self-energy functions $\Pi_{ij}$ and
their derivatives $\Pi^\prime_{ij} = \frac{d \Pi}{d p^2}$, are computed at $p^2=0$.
Various contributions from fermion and scalar loops are summarized as follows:
\begin{itemize}
\item Fermion Loop Contribution
\begin{align}
\Pi_{ij}^{ff}=&-\frac{N_c}{16\pi^2}\bigg[(g_{iL}g_{jL}+g_{iR}g_{jR})\Big(4B_{00}(0,m_{f1}^2,m_{f2}^2) 
-(m_{f1}^2+m_{f2}^2)B_0(0,m_{f1}^2,m_{f2}^2)-A_0(m_{f1}^2)-A_0(m_{f2}^2)\Big)\notag\\
&+2m_{f1}m_{f2}(g_{iL}g_{jR}+g_{iR}g_{jL})B_0(0,m_{f1}^2,m_{f2}^2)\bigg],\\
\Pi_{ij}^{'ff}=&-\frac{N_c}{16\pi^2}\bigg[(g_{iL}g_{jL}+g_{iR}g_{jR})\Big(4B'_{00}(0,m_{f1}^2,m_{f2}^2) 
-(m_{f1}^2+m_{f2}^2)B'_0(m_{f1}^2,m_{f2}^2)+B_0(0,m_{f1}^2,m_{f2}^2)\Big)\notag\\
&+2m_{f1}m_{f2}(g_{iL}g_{jR}+g_{iR}g_{jL})B'_0(0,m_{f1}^2,m_{f2}^2)\bigg],
\end{align}
where $g_{iL}$ and $g_{iR}$ are the left-handed and right-handed 
couplings of the gauge boson labelled $i$ to the fermions running in the loop, from the vertex
\bea
\bar{f}_1 \gamma^\mu (g_{iL}P_L+g_{iR}P_R) f_2 A_\mu.
\eea

\item Scalar Tadpole Contribution
\begin{align}
&\Pi_{ij}^s=-gA_0(m_s^2),\\
&\Pi_{ij}^{'s}=0,
\end{align}
where $g$ is the coupling strength of the $s-s-V-V$ four-point coupling.

\item Scalar Loop Contribution
\begin{align}
&\Pi_{ij}^{ss}=4g_ig_jB_{00}(0,m_{s1}^2,m_{s2}^2),\\
&\Pi_{ij}^{'ss}=4g_ig_jB'_{00}(0,m_{s1}^2,m_{s2}^2),
\end{align}
where $g_i$ is the coupling strength of the $s-s-V$ three-point coupling involving gauge boson labelled $i$.

\item Scalar-Vector Loop Contribution
\begin{align}
&\Pi_{ij}^{vs}=-g_i g_j B_0(0,m_v^2,m_s^2),\\
&\Pi_{ij}^{'vs}=-g_i g_j B'_0(0,m_v^2,m_s^2),
\end{align}
where $g_i$ is the coupling strength of the $s-V-V$ three-point coupling involving gauge boson labelled $i$.

\end{itemize}

\subsection{The Oblique Parameters $S, T$}

We present the contributions to the $S, T$ from the fermion loops and the boson loops, seperately. 

\subsubsection{The Fermion Loops}
The $T$ parameter is computed as
\begin{align}
\alpha T \equiv \frac{e^2}{s_W^2 c_W^2 m_Z^2}\left [\Pi_{11}(0) - \Pi_{33}(0) \right] 
= \frac{\Pi_{WW}}{m_W^2}-\frac{\Pi_{ZZ}}{m_Z^2}-\frac{2s_W}{c_W}\frac{\Pi_{Z\gamma}}{m_Z^2} ,
\end{align}
where for fermion loops the last term on the right-hand side does not contribute. 
In the SM, the fermion contributions mainly come from the third generation quarks:
\begin{align}
&\Pi_{WW}^{SM}=\Pi_{WW}^{tb,SM},\\
&\Pi_{ZZ}^{SM}=\Pi_{ZZ}^{tt,SM}+\Pi_{ZZ}^{bb,SM}.
\end{align}
In our model, there are new contributions from the vector-like fermion $T$:
\begin{align}
\Pi_{WW}&=\Pi_{WW}^{tb}+\Pi_{WW}^{Tb}, \\
\Pi_{ZZ}&=\Pi_{ZZ}^{tt}+2\Pi_{ZZ}^{tT}+\Pi_{ZZ}^{TT}+\Pi_{ZZ}^{bb,SM}.
\end{align}
Subtracting the SM contribution, the NP correction on the parameter $T$ from the fermion loops can be obtained
\begin{equation}
\Delta T_F=\frac{1}{\alpha}\Big[\frac{1}{m_W^2}\left(\Pi_{WW}^{tb}+\Pi_{WW}^{Tb}
-\Pi_{WW}^{tb,SM}\right)-\frac{1}{m_Z^2}\left(\Pi_{ZZ}^{tt}+2\Pi_{ZZ}^{tT}+\Pi_{ZZ}^{TT}-\Pi_{ZZ}^{tt,SM}\right)\Big].
\end{equation}

The $S$ parameter is defined as
\begin{align}
S=16\pi\big(\Pi'_{33}-\Pi'_{3\gamma}\big).
\end{align}
In the SM, the fermion contributions are
\begin{align}
\Pi_{33}^{\prime SM}=\Pi_{33}^{\prime tt,SM}+\Pi_{33}^{\prime bb,SM},\\
\Pi_{3\gamma}^{\prime SM}=\Pi_{3\gamma}^{\prime tt,SM}+\Pi_{3\gamma}^{\prime bb,SM}.
\end{align}
while the new model gives
\begin{align}
\Pi_{33}^{\prime SM}=\Pi_{33}^{\prime tt}+2\Pi_{33}^{\prime tT}+\Pi_{33}^{\prime TT}+\Pi_{33}^{\prime bb,SM},\\
\Pi_{3\gamma}^{\prime SM}=\Pi_{3\gamma}^{\prime tt}+2\Pi_{3\gamma}^{\prime tT}+\Pi_{3\gamma}^{\prime TT}+\Pi_{3\gamma}^{\prime bb,SM}.
\end{align}
Therefore the NP correction on the parameter $S$ from the fermion loops is
\begin{equation}
\Delta S_F=16\pi\left[\Big(\Pi_{33}^{\prime tt}+2\Pi_{33}^{\prime tT}+\Pi_{33}^{\prime TT}-\Pi_{33}^{\prime tt,SM}\Big)
-\Big(\Pi_{3\gamma}^{\prime tt}+2\Pi_{3\gamma}^{\prime tT}+\Pi_{3\gamma}^{\prime TT}-\Pi_{3\gamma}^{\prime tt,SM}\Big)\right].
\end{equation}

\subsubsection{The Boson Loops}

To compute the boson contributions, it is convenient to split them 
into the gauge parts $(\widetilde{T}_v, \widetilde{S}_v)$ and the scalar parts $(\widetilde{T}_s, \widetilde{S}_s)$: 
\begin{equation}
T_S=\widetilde{T}_v+\widetilde{T}_s,\quad S_S=\widetilde{S}_v+\widetilde{S}_s,
\end{equation}
where the tilde indicates there are divergences in each part $(\widetilde{T}, \widetilde{S})$, while the total boson constributions to $(T_S, S_S)$ are convergent. 
The gauge parts consist of the contributions from the $W/Z$ loops, the ghost loops and the Goldstone loops,  
which are not altered in our model. So the gauge parts will not contribute to $\Delta T_S$ and $\Delta S_S$ in our model.
The scalar parts $(\widetilde{T}_s,\widetilde{S}_s)$ consist of all the loops that involve the Higgs boson, 
or any new real scalars that mix with the Higgs boson. In the following, we will only consider the contributions to the $\Delta T_S$ and $\Delta S_S$ from the scalar parts.

In the SM, the self-energy functions involving the Higgs boson are
\begin{align}
&\Pi_{WW}^{SM}=\frac{1}{2}\Pi_{WW}^{h,SM}+\Pi_{WW}^{hW,SM}+\Pi_{WW}^{h\pi^{\pm},SM},\\
&\Pi_{ZZ}^{SM}=\frac{1}{2}\Pi_{ZZ}^{h,SM}+\Pi_{ZZ}^{hZ,SM}+\Pi_{ZZ}^{h\pi^0,SM},
\end{align}
while there is no Higgs boson contribution in the two-point function $\Pi_{Z\gamma}$ . 
Inserting the above self-energy functions back to $T$ parameter definition, 
due to the cancellation between the first terms in the $W$ and $Z$ self-energy functions, we obtain the scalar part $\widetilde{T}_s^{SM}$
\begin{align}
\widetilde{T}_s^{SM}&=\frac{1}{\alpha}\Big[\frac{\Pi_{WW}^{hW,SM}+\Pi_{WW}^{h\pi^{\pm},SM}}{m_W^2}-\frac{\Pi_{ZZ}^{hZ,SM}+\Pi_{ZZ}^{h\pi^0,SM}}{m_Z^2}\Big],\\
&=\frac{3}{16\pi c_W^2}\Delta+T_s(m_h),
\label{eq:appTs}
\end{align}
where $\Delta$ is the divergent term $\Delta = \frac{2}{4-d} - \gamma + \ln 4\pi \mu^2$ in the $\overline{MS}$ scheme. 
Here
the finite part is written as a function of the scalar mass $m$ 
\begin{align}
\label{eq:FunTs}
T_s(m)=-\frac{3}{16\pi c_W^2}\left[\frac{1}{(m^2-m_Z^2)(m^2-m_W^2)}\left(m^4\ln m^2-s_W^{-2}(m^2-m_W^2)m_Z^2\ln m_Z^2+s_W^{-2}c_W^2(m^2-m_Z^2)m_W^2\ln m_W^2\right)-\frac{5}{6}\right].
\end{align}
The calculation is done in the $\overline{MS}$ scheme, without loss of generality.

In our model, the self-energy functions involving the Higgs boson and the new scalar are 
\begin{align}
&\Pi_{WW}=\frac{1}{2}(\Pi_{WW}^{h}+\Pi_{WW}^{S})+(\Pi_{WW}^{hW}+\Pi_{WW}^{SW})+(\Pi_{WW}^{h\pi^{\pm}}+\Pi_{WW}^{S\pi^{\pm}}),\\
&\Pi_{ZZ}=\frac{1}{2}(\Pi_{ZZ}^{h}+\Pi_{ZZ}^{S})+(\Pi_{ZZ}^{hZ}+\Pi_{ZZ}^{SZ})+(\Pi_{ZZ}^{h\pi^0}+\Pi_{ZZ}^{h\pi^0}).
\end{align}
Note that the first terms do not contribute to the $T$ parameter, as in the SM. 
Similarly, we obtain the scalar part $\widetilde{T}_s$
\begin{align}
\widetilde{T}_s&=\frac{1}{\alpha}\Big[\frac{\Pi_{WW}^{hW}+\Pi_{WW}^{h\pi^{\pm}}}{m_W^2}-\frac{\Pi_{ZZ}^{hZ}+\Pi_{ZZ}^{h\pi^0}}{m_Z^2}\Big]
+\frac{1}{\alpha}\Big[\frac{\Pi_{WW}^{SW}+\Pi_{WW}^{S\pi^{\pm}}}{m_W^2}-\frac{\Pi_{ZZ}^{SZ}+\Pi_{ZZ}^{S\pi^0}}{m_Z^2}\Big]\\
&=c_{\varphi}^2\left(\frac{3\Delta}{16\pi c_W^2}+T_s(m_h)\right)+s_{\varphi}^2\left(\frac{3\Delta}{16\pi c_W^2}+T_s(m_S)\right)\\
&=\frac{3\Delta}{16\pi c_W^2}+c_{\varphi}^2T_s(m_h)+s_{\varphi}^2T_s(m_S),
\end{align}
where the divergent part is the same as the SM as expected.
As the new scalar only contribute to the part $\widetilde{T}_s$ via mixing with the SM Higgs,
common factors $s_\varphi$ and $c_\varphi$ could be extracted,
leaving a SM-like $\tilde{T}_s(m)$.
Hence the function  $T_s(m)$ can be used to obtain a concise form for the $\widetilde{T}_s$ in our model.
Subtracting the SM contribution $\widetilde{T}_s^{SM}$ in Eq.~\ref{eq:appTs}, we get a finite and very concise result for $\Delta T_S$
\begin{equation}
\Delta T_S=\Delta \widetilde{T}_s=s_{\varphi}^2\Big[T_s(m_S)-T_s(m_h)\Big],
\end{equation}

%


The $S$ parameter can be defined in an alternative way  using the hyperchage $Y$, 
\begin{equation}
S \equiv -16\pi\Pi'_{3Y},
\end{equation}
which  reduces a lot of work for boson contributions. 
The Higgs dependent part $\widetilde{S}_s$ in the SM is
\begin{align}
\widetilde{S}_s^{SM}&=-16\pi\Big(\Pi_{3Y}^{\prime hZ,SM}+\Pi_{3Y}^{\prime h\pi^0,SM}\Big)\\
&=-\frac{1}{12\pi}\Delta+S_s(m_h),
\end{align}
where
\begin{align}
\label{eq:FunSs}
S_s(m)&=\frac{1}{12\pi}\left[\ln m^2-\frac{5}{6}-\frac{(4m^2+6m_Z^2)m_Z^2}{(m^2-m_Z^2)^2}+\frac{(9m^2+m_Z^2)m_Z^4}{(m^2-m_Z^2)^3}\ln\frac{m^2}{m_Z^2}\right].
\end{align}
For the same reason as in $T_S$ calculation, when we turn to the new model, it has a very concise form
\begin{align}
\widetilde{S}_s&=c_{\varphi}^2\Big(-\frac{\Delta}{12\pi}+S_s(m_h)\Big)+s_{\varphi}^2\Big(-\frac{\Delta}{12\pi}+S_s(m_S)\Big)\\
&=-\frac{\Delta}{12\pi}+c_{\varphi}^2S_s(m_h)+s_{\varphi}^2S_s(m_S),
\end{align}
and similarly
\begin{align}
\Delta S_S=\Delta \widetilde{S}_s=s_{\varphi}^2\Big[S_s(m_S)-S_s(m_h)\Big].
\end{align}

\section{The Partial Decay Widths of the Heavy Particles}

The heavy scalar $S$ mainly decays in the following channels with partial widths:
\begin{align}
\Gamma(S\to ZZ)&=\frac{G_Fm_S^3}{16\sqrt{2}\pi}s_{\varphi}^2
\times\lambda(1,r_Z,r_Z)^{\frac{1}{2}}(1-4r_Z+12r_Z^2),\\
\Gamma(S\to WW)&=\frac{G_Fm_S^3}{8\sqrt{2}\pi}s_{\varphi}^2
\times\lambda(1,r_Z,r_Z)^{\frac{1}{2}}(1-4r_W+12r_W^2),\\
\Gamma(S\to hh)&=\frac{G_Fm_S^3}{16\sqrt{2}\pi}s_{\varphi}^2c_{\varphi}^2(c_\varphi+\frac{v}{u}s_\varphi)^2
\times\lambda(1,r_h,r_h)^{\frac{1}{2}}(1+2r_h)^2,\\
\Gamma(S\to Tt)&=\frac{3G_Fm_Sm_T^2}{8\sqrt{2}\pi}s_L^2c_L^2(s_{\varphi}-\frac{v}{u}c_{\varphi})^2
\times\lambda(1,r_T,r_t)^{\frac{1}{2}}(1+\frac{r_t}{r_T})\Big[1-(r_T-r_t)^2\Big],\\
\Gamma(S\to tt)&=\frac{3G_Fm_Sm_t^2}{4\sqrt{2}\pi}(s_{\varphi}c_L^2+\frac{v}{u}c_{\varphi}s_L^2)^2
\times\lambda(1,r_t,r_t)^{\frac{3}{2}},
\end{align}
where $r_X=m_X^2/m_S^2$ and the kinematic function $\lambda$ is
\begin{align}
\lambda(1,r_1,r_2)&=1+r_1^2+r_2^2-2r_1-2r_2-2r_1r_2. 
\end{align}
Note that if $\tan\varphi\sim \frac{v}{u}$, the partial width in the $Tt$ channel is suppressed,
while if $\tan\varphi\sim -\frac{v}{u}\tan^2\theta_L$, the $tt$ channel is suppressed. 
%

The vector-like fermion mainly decays in the following channels with partial widths:
\begin{align}
\Gamma(T\to Zt)&=\frac{G_Fm_T^3}{16\sqrt{2}\pi}s_L^2c_L^2
\times \lambda(1,r_Z,r_t)^{\frac{1}{2}}(1+r_Z-2r_t-2r_Z^2+r_tr_Z+r_t^2),\\
\Gamma(T\to Wb)&=\frac{G_Fm_T^3}{8\sqrt{2}\pi}s_L^2
\times \lambda(1,r_W,r_b)^{\frac{1}{2}}(1+r_W-2r_b-2r_W^2+r_br_W+r_b^2),\\
\Gamma(T\to ht)&=\frac{G_Fm_T^3}{16\sqrt{2}\pi}s_L^2c_L^2(c_{\varphi}+\frac{v}{u}s_{\varphi})^2
\times \lambda(1,r_h,r_t)^{\frac{1}{2}}(1+6r_t-r_h-r_tr_h+r_t^2),\\
\Gamma(T\to St)&=\frac{G_Fm_T^3}{16\sqrt{2}\pi}s_L^2c_L^2(s_{\varphi}-\frac{v}{u}c_{\varphi})^2
\times\lambda(1,r_S,r_t)^{\frac{1}{2}}(1+6r_t-r_S-r_tr_S+r_t^2).
\end{align}
Also note that the $St$ channel is suppressed due to the factor $s_{\varphi}-\frac{v}{u}c_{\varphi}$ in its coupling. 

\end{widetext}



\end{document}